\documentclass[%
 reprint,
preprintnumbers,
 amsmath,amssymb,
 aps,
]{revtex4-2}

\usepackage{graphicx}% Include figure files
\usepackage{dcolumn}% Align table columns on decimal point
\usepackage{bm}% bold math
\usepackage{hyperref}% add hypertext capabilities
\usepackage[T1]{fontenc}
\usepackage{xcolor}

\usepackage{tikz}
\usepackage{amsmath}
\usepackage{bm}
\usepackage{soul}
\usepackage{xspace}

\newcommand{\eps}{{\varepsilon}}
\renewcommand{\O}{{\mathcal O}}

\newcommand{\oneloop}{
\begin{minipage}{2.5cm}
\begin{tikzpicture}
\draw[ thick, -] (0,0.5) arc (90:450:0.5);
\draw[ thick, -] (-1,0) -- (-0.5,0);
\draw[ thick, -] (0.5,0) -- (0.85,0.35);
\draw[ thick, -] (0.5,0) -- (0.85,-0.35);
\filldraw (-0.5,0) circle (2pt);
\filldraw (0.5,0) circle (2pt);
\draw (-1.2,0.0) node{${\scriptstyle {p_1}}$};
\draw (1.05,0.55) node{${\scriptstyle {p_2}}$};
\draw (1.05,-0.55) node{${\scriptstyle {p_3}}$};
\end{tikzpicture}
\end{minipage}}

\newcommand{\oneloopa}{
\begin{minipage}{2.5cm}
\begin{tikzpicture}
\draw[ thick, -] (0,0.5) arc (90:450:0.5);
\draw[ thick, -] (-1,0) -- (-0.5,0);
\draw[ thick, -] (0.5,0) -- (1,0);
\filldraw (-0.5,0) circle (2pt);
\filldraw (0.5,0) circle (2pt);
\draw (-1.2,0.0) node{${\scriptstyle {p_1}}$};
\draw (1.2,0.0) node{${\scriptstyle {p_2}}$};
\end{tikzpicture}
\end{minipage}}

\newcommand{\twoloop}{
\begin{minipage}{3.2cm}
\begin{tikzpicture}
\draw[ thick, -] (-1,0) -- (0.5,0.75);
\draw[ thick, -] (-1,0) -- (0.5,-0.75);
\draw[ thick, -] (0.5,-0.75) .. controls (0,-0.3) and (0,0.3).. (0.5,0.75);
\draw[ thick, -] (0.5,-0.75) .. controls (1,-0.3) and (1,0.3) .. (0.5,0.75);
\draw[ thick, -] (-1,0) -- (-1.5,0);
\draw[ thick, -] (1,-1.25) -- (0.5,-0.75);
\draw[ thick, -] (1,1.25) -- (0.5,0.75);
\filldraw (-1,0) circle (2pt);
\filldraw (0.5,0.75) circle (2pt);
\filldraw (0.5,-0.75) circle (2pt);
\draw (-1.7,0.0) node{${\scriptstyle {p_3}}$};
\draw (1.2,1.25) node{${\scriptstyle {p_1}}$};
\draw (1.2,-1.25) node{${\scriptstyle {p_2}}$};
\end{tikzpicture}
\end{minipage}}

\newcommand{\twoloopa}{
\begin{minipage}{2.5cm}
\begin{tikzpicture}
\draw[ thick, -] (-1,0) -- (0.5,0.75);
\draw[ thick, -] (-1,0) -- (0.5,-0.75);
\draw[ thick, -] (0.5,-0.75) .. controls (0,-0.3) and (0,0.3).. (0.5,0.75);
\draw[ thick, -] (0.5,-0.75) .. controls (1,-0.3) and (1,0.3) .. (0.5,0.75);
\draw[ thick, -] (1,-1.25) -- (0.5,-0.75);
\draw[ thick, -] (1,1.25) -- (0.5,0.75);
\filldraw (-1,0) circle (2pt);
\filldraw (0.5,0.75) circle (2pt);
\filldraw (0.5,-0.75) circle (2pt);
\draw (1.2,1.25) node{${\scriptstyle {p_1}}$};
\draw (1.2,-1.25) node{${\scriptstyle {p_2}}$};
\end{tikzpicture}
\end{minipage}}

\newcommand{\gsuba}{
\begin{minipage}{2.6cm}
\begin{tikzpicture}
\draw[ thick, -] (0,0.5) arc (90:450:0.5);
\draw[ thick, -] (-1,0) -- (-0.5,0);
\draw[ thick, -] (0.5,0) -- (0.85,0.35);
\draw[ thick, -] (0.5,0) -- (0.85,-0.35);
\filldraw (-0.5,0) circle (2pt);
\filldraw (0.5,0) circle (2pt);
\draw (-1.2,0.0) node{${\scriptstyle {p_3}}$};
\draw (1.05,0.55) node{${\scriptstyle {p_1}}$};
\draw (1.05,-0.55) node{${\scriptstyle {p_2}}$};
\end{tikzpicture}
\end{minipage}}

\newcommand{\gsubb}{
\begin{minipage}{1.2cm}
\begin{tikzpicture}
\draw[ thick, -] (0.5,-0.75) .. controls (0,-0.3) and (0,0.3).. (0.5,0.75);
\draw[ thick, -] (0.5,-0.75) .. controls (1,-0.3) and (1,0.3) .. (0.5,0.75);
\draw[ thick, -] (0.5,-1.25) -- (0.5,-0.75);
\draw[ thick, -] (0.5,1.25) -- (0.5,0.75);
\filldraw (0.5,0.75) circle (2pt);
\filldraw (0.5,-0.75) circle (2pt);
\draw (0.7,1.25) node{${\scriptstyle {p_1}}$};
\draw (0.7,-1.25) node{${\scriptstyle {p_2}}$};
\end{tikzpicture}
\end{minipage}}

\begin{document}

\preprint{CERN-TH-2025-135}

\title{Multi-loop spectra in general scalar EFTs and CFTs}

\author{Johan Henriksson$^{a}$}
\author{Franz Herzog$^b$}
\author{Stefanos R. Kousvos$^{cd}$}
\author{Jasper Roosmale Nepveu$^{ef}$}

\affiliation{$^{a}$Theoretical Physics Department, CERN, 1211, Geneva, Switzerland}

\affiliation{$^{b}$Higgs Centre for Theoretical Physics, School of Physics and Astronomy,\\The University of Edinburgh, Edinburgh EH9 3FD, Scotland, UK}

\affiliation{$^{c}$Department of Physics, University of Pisa and INFN, section of Pisa, Largo Pontecorvo 3, I-56127 Pisa, Italy}
\affiliation{$^{d}$Department of Physics, University of Torino and INFN, section of Torino, Via P.\ Giuria 1, 10125 Torino, Italy}

\affiliation{$^{e}$Department of Physics and Center for Theoretical Physics, National Taiwan University, Taipei 10617, Taiwan}
\affiliation{$^{f}$Leung Center for Cosmology and Particle Astrophysics, Taipei 10617, Taiwan}

\begin{abstract}
We consider the most general effective field theory (EFT) Lagrangian with scalar fields and derivatives, and renormalise it to substantially higher loop order than existing results in the literature. EFT Lagrangians have phenomenological applications, for example by encoding corrections to the Standard Model from unknown new physics. At the same time, scalar EFTs capture the spectrum of Wilson--Fisher conformal field theories (CFTs) in $4-\varepsilon$ dimensions. Our results are enabled by a more efficient version of the $R^*$ method for renormalisation, in which the IR divergences are subtracted via a small-momentum asymptotic expansion. In particular, we renormalise the most general set of composite operators up to engineering dimension six and Lorentz rank two. We exhibit direct applications of our results to Ising ($Z_2$), $O(n)$, and hypercubic ($S_n \ltimes (Z_2)^n$) CFTs, relevant for a plethora of real-world critical phenomena. The computed scaling dimensions agree well with known non-perturbative results, and they lead to new predictions where such results do not yet exist. We thereby expand the understanding of generic EFTs and open new possibilities in diverse fields, such as the numerical conformal bootstrap.
\end{abstract}

\maketitle

\section{Introduction}

A remarkable feature of quantum field theory is its range of applications: from the search for new physics at energy scales above 1~TeV to emergent low-energy conformal field theories describing phase transitions in real-world fluids and crystals. Here we will show that these two disparate regimes not only fall under the same paradigm, but also benefit from the same concrete computational advances.

Consider a renormalisable Lagrangian in $d$ spacetime dimensions perturbed by higher-dimensional operators,
\begin{equation}
\label{eq:lagr1}
\mathcal L=\mathcal L_{\text{renorm.}}+\sum_ic_i \O_i\,.
\end{equation}
This defines a low energy effective field theory (EFT). The couplings $c_i$ scale naively with the inverse powers of the scale of new physics, $c_i\sim M^{d-[\O_i]}$, where $[\O_i]$ denotes the engineering dimension of $\O_i$. The naive scaling receives quantum corrections, 
leading to operator mixing and running determined by the anomalous dimensions. 
In e.g.~the Standard Model (SM) EFT, such effects are under systematic study, leading to standardised operator bases \cite{Grzadkowski:2010es,Murphy:2020rsh} and multi-loop renormalisation results~\cite{deVries:2019nsu,Aebischer:2022anv,Jenkins:2023bls,Ibarra:2024tpt,DiNoi:2024ajj,Born:2024mgz,Duhr:2025zqw,Zhang:2025ywe}.

Conformal field theories (CFTs) \cite{Ferrara:1973yt,Osborn:1993cr,Rychkov:2016iqz,Simmons-Duffin:2016gjk} are believed to describe continuous phase transitions across classical and quantum critical phenomena \cite{Pelissetto:2000ek}. In many cases they can be realised as IR fixed-points of quantum field theories, either by a long uncontrolled renormalisation group flow or in perturbative limits such as the $\eps$-expansion \cite{Wilson:1971dc,Wilson:1973jj}. The spectrum of the IR CFT can then be extracted by computing anomalous dimensions of primary operators. 
In EFT, a global symmetry is generally assumed, and only Lorentz scalar operators singlet under this symmetry are included in the sum in~\eqref{eq:lagr1}. 
However, to describe the full spectrum of an IR CFT, also non-singlets and non-scalar operators need to be considered (the $c_i$ are viewed as probes). In particular, the modern axiomatic/bootstrap approach to CFT emphasises the set of local operators and their associated data as a defining property of the theory, where scaling dimensions (classical plus anomalous) are key pieces of data.

\begin{figure}[t]\hspace{-6mm}
\includegraphics[scale=0.61]{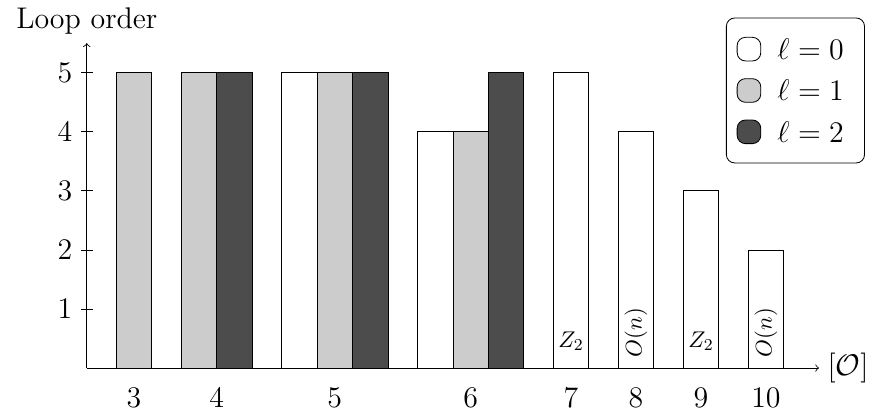}
\caption{
Overview of our results for all operators in the general theory at spin $\ell$ and mass dimension $[\mathcal{O}]$. Beyond dimension six, we provide results for the single scalar EFT ($Z_2$) and the $O(n)$ model (singlets only).
}\label{fig:summary}
\end{figure}

Here we shall renormalise the most general scalar field theory in $d=4-\eps$ dimensions in the (modified) minimal subtraction ($\overline{\text{MS}}$) scheme. 
The renormalizable part at zero spin,
\begin{eqnarray}
\nonumber
&\mathcal{L}_\text{renorm.} = \frac12(\partial \phi)^2-\Lambda-t_a\phi^a
- \frac12m^2_{ab}\phi^a\phi^b\\&
-\, \frac1{3!}h_{abc}\,\phi^a\phi^b\phi^c- \frac{1}{4!}\lambda_{abcd}\phi^a\phi^b\phi^c\phi^d    \,,
\end{eqnarray}
was renormalised to six-loop order in \cite{Bednyakov:2021ojn}, and to seven-loop order in \cite{Schnetz:2022nsc} under the $O(n)$ symmetric restriction with $\Lambda=h_{abc}=0$. 
We make the conceptual and technical leap to the systematic treatment of higher-dimensional operators $c_i\O_i$ at multi-loop order in this general theory, as well as spinning RG-relevant operators. 
Going beyond the typical EFT restriction to singlet and scalar operators, we consider all operators up to engineering dimension six and Lorentz rank two, 
\begin{equation}
    \mathcal{L} = \frac12 (\partial\phi)^2 - \frac1{4!}\lambda_{abcd}\, \phi^a\phi^b\phi^c\phi^d 
    + \sum_{\ell=0}^2% =0,1,2}
      \sum_{\Delta=2}^6 %=\ell+2}
      c^{(\Delta,\ell)}_i \mathcal{O}^{(\Delta,\ell)}_i\,.
\end{equation}
For example, four-field operators at spin-0 and spin-1 are
\begin{align}\label{eq:exO}
    c^{(6,0)}\O^{(6,0)} &= c_{abcd}^{(6,0)}\, \phi^a\phi^b\partial_\mu\phi^c\partial^\mu \phi^d\,,\\
    c^{(6,1)}\O^{(6,1)} &= u^\mu\, c_{abcd}^{(6,1)}\, \phi^a\phi^b\phi^c\partial_\mu \phi^d\,,
\end{align}
where $u^\mu$ is a reference vector.
We provide the full Lagrangian in the supplemental material.
For $Z_2$ and $O(n)$ symmetry, we extend the analysis further; see Fig.~\ref{fig:summary} for a visual summary. We restrict to single insertions of the higher-dimensional operator, sufficient to compute anomalous dimensions.

We will work with a specific \emph{primary operator basis} to systematically identify non-redundant degrees of freedom. 
This basis is well-suited for subsequent treatment in both EFT and CFT. 
We will compute anomalous dimensions to dramatically increased loop order compared to the literature. For CFT applications, the data is re-expanded as a series in $\eps=4-d$ using the critical coupling. Typically we extend existing $O(\eps)$ results to new estimates at $O(\eps^5)$. The power of these estimates is demonstrated by comparing with non-perturbative results for three-dimensional CFTs ($\eps=1$), as shown in the figures and tables below.
In these comparisons we use Pad\'e$_{m,n}$ approximants, which are rational functions $(a_0+a_1\eps+\ldots+ a_m\eps^m)/(1+b_1\eps+\ldots+b_n\eps^n)$ that reproduce the perturbative series up to order $\eps^{m+n}$. We take $m\leq n\leq m+1$ unless otherwise stated.

\section{Method}

\subsection{Primary operator basis}

EFT Lagrangians are plagued by redundancies. For instance, integration by parts (at the action level) and field redefinitions alter the form of the Lagrangian without affecting the S-matrix. 
Since the set of independent operators that contribute to the EFT S-matrix coincides with the set of primary operators in CFT~\cite{Henning:2015alf,Henning:2017fpj}, we derive a minimal basis by imposing the primary condition,
\begin{equation}\label{eq:primaryCondition}
    [K_\mu,c_A^{(\Delta,\ell)}\mathcal{O}^{(\Delta,\ell)}_A(0)]=0 
    \ \leftrightarrow \ 
    c^{(\Delta,\ell)}_A\mathcal{O}^{(\Delta,\ell)}_A(x)\subset\mathcal{L}\,,
\end{equation}
where $K_\mu$ is the generator of special conformal transformations and {\small $A$} stands for a collection of flavour indices.
The primary condition imposes constraints on the coupling constant tensors.
For instance, the coefficient of the operator in \eqref{eq:exO} must satisfy
\begin{equation}\label{eq:cond6}
    c_{abcd}^{(6,0)} + c_{bcad}^{(6,0)} 
    + c_{cabd}^{(6,0)} = 0\,.
\end{equation}
This removes all redundant parameters from the tensor. 
We call this basis of operators the \emph{primary operator basis}.

Notably, in the primary basis we do not include operators proportional to $\partial^2\phi$. However, they are generated as counterterms and we take their effect into account. Computationally, at linear order in the couplings (except for $\lambda_{abcd}$), we
subtract all operators proportional to the interacting equations of motion, $\partial^2 \phi^a + \frac16\lambda_{abcd}\phi^b\phi^c\phi^d = 0$.
We follow the setup outlined in~\cite{Cao:2021cdt} for this procedure. 
In particular, by automatically subtracting subdivergences from Feynman integrals, we avoid the need for explicit counterterm graphs. 
In comparison to~\cite{Cao:2021cdt}, we have considerably improved on the computation of UV divergences, as we will now describe.

\subsection{Framework for computing UV counterterms}

We extract the renormalisation constants from 
correlation functions of $n$ fields $\phi^a$
% $n$-point correlation functions 
with operator insertions using a new formulation of the $R^*$ method \cite{Vladimirov:1979zm,Chetyrkin:1982nn,Chetyrkin:1984xa,Smirnov:1985yck,Larin:2002sc,Kleinert:2001hn,Chetyrkin:2017ppe,Herzog:2017bjx,deVries:2019nsu,Beekveldt:2020kzk}, first presented in \cite{Chakraborty2023}. The key advantage of the $R^*$ approach is that it allows to extract the renormalisation constants of $L$-loop $n$-point correlators from Feynman integrals of at most $L-1$ loops. 
This reduction in complexity is achieved by a suitable infrared rearrangement (IRR) that reroutes the external momenta in a diagram in such a way that the integration over loop momenta can be factorised~\cite{Chetyrkin:1982nn}. 

The main difference between the original formulation of $R^*$ and our implementation is in the way infrared divergences, which may arise due to IRR, are subtracted. Rather than using the local IR subtraction operation, we use a small-momentum asymptotic expansion as implemented via the expansion-by-subgraph \cite{Chetyrkin:1982zq, Chetyrkin:1983qlc, Gorishnii:1983su, Gorishnii:1986mv, LlewellynSmith:1987jx, Gorishnii:1986gn, Chetyrkin:1988zz, Chetyrkin:1988cu, Gorishnii:1989dd, Smirnov:1990rz, Smirnov:1994tg,Smirnov:2002pj,Chakraborty:2024uzz}. We will refer to this variant of $R^*$ as $R^*_{\text{ME}}$. In comparison to the previous approach used by one of the authors~\cite{Herzog:2017bjx,deVries:2019nsu}, $R^*_{\text{ME}}$ leads to far less counterterms, especially in the context of higher-dimensional operators.  

We start with Bogoliubov's recursive definition (BPHZ) \cite{Bogoliubov:1957gp,Hepp:1966eg,Zimmermann:1969jj} of the local UV counterterm $\mathcal{Z}(G)$ of a Feynman graph $G$ in the MS (or $\overline{\text{MS}}$) scheme~\cite{Caswell:1981ek}, 
\begin{equation}
\mathcal{Z}(G)=-K_\eps \bar R (G),\quad    \bar R (G)=\sum_{\gamma \subsetneq G } \mathcal{Z}(\gamma) *G/\gamma \,,  
\end{equation}
where the sum goes all UV-divergent bridgeless proper subgraphs, $\gamma=\sqcup \gamma_i$, of $G$.  $G/\gamma$ denotes the contracted graph obtained after contracting each connected component $\gamma_i$ of $\gamma$ into a vertex in $G$, into which the counterterm $\mathcal{Z}(\gamma)=\prod_i \mathcal{Z}(\gamma_i)$ is inserted, as indicated by the $*$-symbol. 
$K_\eps$ implements integration and projects out pole terms in $\eps$. 

Since $\mathcal{Z}(G)$ is polynomial (local) in the external momenta of degree $\omega(G)$, the superficial degree of divergence of $G$, the counterterm can be written as
\begin{align}
\mathcal{Z}(G)&=T^{(\omega)}_{\{p\}}\mathcal{Z}(G') =-K_\eps   \widetilde {T}^{(\omega)}_{\{p\}}  \bar R (G') \,.  
\end{align}
Here $G'$ is the IRR version of $G$ obtained by inserting an arbitrary momentum flow $Q$ in and out of the diagram $G$. The original counterterm of $G$ is recovered by projecting out the polynomial terms of the original external momenta $\{p\}$ in the diagram via the degree-$\omega$ Taylor expansion operator $T^{(\omega)}_{\{p\}}$. 
Importantly, $T^{(\omega)}_{\{p\}}$ does not commute with integration. 
Instead, the corresponding asymptotic expansion operator $\widetilde {T}^{(\omega)}_{\{p\}}$ does and thus can act on the integrand before integration. 
More precisely, it is defined via the expansion-by-subgraph as follows:
\begin{equation}
\widetilde {T}^{(\omega)}_{\{p\}}(G)=\sum_{\gamma_A \subset G} T^{(\omega)}_{\{p\}}(\gamma_A)* G/\gamma_A\,,
\end{equation}
where $\gamma_A$ is an asymptotically irreducible (AI) subgraph; see ref.~\cite{Smirnov:2002pj} for a precise definition. For massless graphs it is sufficient that the AI subgraph contains all hard external legs and that it becomes one-vertex-irreducible when these external lines are connected at an additional vertex. The vertices for $Q$ going in and out of the diagram can be chosen such that the integration of the loop over $G'$ can be factorized, leading to the aforementioned simplification of the IRR, i.e.~the $L$-loop multi-scale integral effectively factorizes into a product of massless propagator-type integrals of at most $L-1$ loops. An example is presented in the supplementary material. 

We have implemented $R^*_{\text{ME}}$ in a program based on \texttt{Maple} \cite{maple} and \texttt{Form}~\cite{Vermaseren:2000nd,Ruijl:2017dtg}. The resulting massless propagator-type integrals, which were computed up to 4 loops in \cite{Baikov:2010hf,Lee:2011jt}, are evaluated using \texttt{Forcer}~\cite{Ruijl:2017cxj}. Tensor reduction is performed with \texttt{Opiter}~\cite{Goode:2024cfy}.
Feynman diagrams were generated with the algorithm of T. Kaneko \cite{Kaneko:1994fd,Kaneko:2017wzd} implemented in \texttt{Form5.0}.

\section{Results and Applications}

Our complete set of data is shared at the \texttt{GitHub} repository
\href{https://github.com/jasperrn/EFT-RGE}{https://github.com/jasperrn/EFT-RGE},
where we also list our conventions. 
We now explain our results and apply them to specific theories. 

The results are given with general field indices. 
For a chosen global symmetry, they first need to be contracted with tensor structures and the coupling constant needs to be substituted. For instance, the $O(n)$-symmetric coupling is $\lambda_{abcd}=\frac\lambda3(\delta_{ab}\delta_{cd}+\delta_{ac}\delta_{bd}+\delta_{ad}\delta_{bc})$. The second step, for CFT applications, is to evaluate the anomalous dimensions at the critical coupling $\lambda_\star$, which is a zero of the $\beta$ function given as a series in $\eps=4-d$. Depending on the symmetry, there may exist none, one, or multiple compatible fixed points. Here we specialize our results to $Z_2$ (Ising CFT), $O(n)$ and cubic symmetry; for which previous results were collected in \cite{Henriksson:2022rnm,Bednyakov:2023lfj}.
Other symmetries are also experimentally relevant \cite{Pelissetto:2000ek}, and can be extracted from our results. 

\subsection{Extraction using tensors}

Let us consider the results for dimension-six scalars. They extend to four-loop order and take the form
\begin{align}
\nonumber
    &\texttt{res[6,0,4,1]}=
 l\bigg(\frac23D^{(6,4)}_{a_4a_3b_1b_2}\lambda_{a_1a_2b_1b_2}+\ldots\bigg)\!+\ldots
   \nonumber\\
   &\texttt{res[6,0,6,1]}=
   l\bigg(\!\frac13D^{(6,4)}_{a_5a_6b_2b_3}\lambda_{a_1b_1b_2b_3}\lambda_{a_2a_3a_4b_1}
   \nonumber\\&
   \qquad\quad  
   +
   D^{(6,6)}_{a_3a_4a_5a_6b_1b_2}\!\lambda_{a_1a_2b_1b_2}+\ldots\bigg)\!+\ldots,
      \label{tensorstruct}
\end{align}
where $l$ is a loop-counting parameter. (We suppressed the classical term at $l^0$.) These are our largest files (40 MB). 
In the repository we also explain the conditions that the coupling constant tensors $D^{(6,i)}$ satisfy. For instance, to extract $O(n)$ singlets we use $D^{(6,4)}_{abcd}=d_4(\delta_{ab}\delta_{cd}-\frac12\delta_{ac}\delta_{bd}-\frac12\delta_{ad}\delta_{bc})$, and $D^{(6,6)}_{abcdef}=d_6(\delta_{ab}\delta_{cd}\delta_{ef}+\text{perms})$, compatible with \eqref{eq:cond6} and total symmetry respectively. 
The coefficients of $d_4$ and $d_6$ in the RHS of \eqref{tensorstruct} (modulo the tensor structures) define the matrix 
\begin{equation}
    \Gamma=\left(\begin{smallmatrix}
        -2\eps+\frac{2n+4}3\lambda-\frac{17n+34}{18}\lambda^2+\ldots\!\!\!
        &
\frac{5(n+2) (n+4)}{81}  \lambda ^3
        \\
        \frac{10(1-n)}9\lambda^3+\ldots &\!\!\!
        -3\eps+(n+14)\lambda+\frac{49n+378}{6}\lambda^2\ldots
    \end{smallmatrix}\right)\,,
    \label{eq:singletsDim6}
\end{equation}
where the top-right corner represents four-loop order, while the bottom left starts at two loops. The full dimension is $\Delta=6+\text{eigs}(\Gamma)$, and in this particular case we can check that the results agree with \cite{Derkachov:1997gc, Jenkins:2023bls,Cao:2021cdt,RoosmaleNepveu:2024zlz}. 

For a non-singlet $O(n)$ representation, consider the $[2,2]$ Young tableau (called $B_4$ in \cite{Henriksson:2022rnm}). Only the $D^{(6,4)}$ term produces a non-zero operator, yielding a one-dimensional entry (i.e.~no mixing matrix):
\begin{align}
    \gamma&\!=\!-2\eps\!+\!2\lambda\!-\!\tfrac{7n+44}{18}\lambda^2\!+\!\left(\tfrac{1516+346 n-11 n^2}{216}\!+\!\tfrac{4n+20}9\zeta_3\right)\lambda^3\nonumber\\&\quad +\ldots.
    \label{eq:B4dim6}
\end{align}
Evaluating \eqref{eq:singletsDim6} and \eqref{eq:B4dim6} at $n=4$ in fact gives results for the three singlet operators of the Higgs sector in the SMEFT. 
That is because the scalar sector of the SM is invariant under $O(4)$ custodial symmetry, and the custodial-violating operator in the $B_4$ representation is invariant under $SU(2)\times U(1)$.
We thereby extend the results in \cite{Jenkins:2023bls} to five loops~\cite{LongPaper}. 
Evaluating them instead at the critical coupling, $\lambda_\star=\frac{3\eps}{n+8}+\frac{9(3n+14)\eps^2}{(n+8)^3}+\ldots$, gives results for operators in the $O(n)$ Wilson--Fisher CFT, where the full scaling dimension is $6+\gamma$.

\subsection{Spectrum of the 3d Ising CFT}

The Ising CFT, the canonical Wilson--Fisher fixed-point, is of central experimental and theoretical importance and has been studied with a variety of methods. In table~\ref{tab:newIsing}, we compare our new results up to $\Delta=6$ with the most precise non-perturbative results from the conformal bootstrap. 

\begin{table}
\caption{Our results at $\Delta\leq6$ in the Ising CFT, compared with numerical bootstrap results in 3d with statistical and rigorous error intervals respectively.}
\label{tab:newIsing}
\begin{tabular}{lllll}
\hline\hline
Operator & $\phi^5$ & $\phi^2\partial_\mu\partial_\nu \phi$  & $\phi^3\partial_\mu\partial_\nu \phi$ 
\\\hline
Previous loop order & $\eps^3$ \cite{Zhang1982} & $\eps^2$ \cite{Bertucci:2022ptt} & $\eps^1$ \cite{Kehrein:1994ff}
\\
New loop order & $\eps^5$ & $\eps^5$ & $\eps^5$
\\
Pad\'e approximant & $5.257395$ & $4.162978$ & $5.465027$
\\
Statistical error \cite{Simmons-Duffin:2016wlq} & $5.2906(11)$  & $4.180305(18)$ &  $5.50915(44) $
\\
Rigorous error \cite{Reehorst:2021hmp} & $5.262(89)$ & --- & $5.499(17) $
\\\hline\hline
\end{tabular}
\end{table}

In this case, our results also include Lorentz scalars with engineering dimension $[\mathcal{O}]\leq10$, which read
\allowdisplaybreaks
\begin{align}
&\Delta_{\phi^5}=5+\tfrac56 \eps  -\tfrac{685}{324} \eps  ^2+6.64393 \eps 
   ^3
   \nonumber\displaybreak[0]\\&\hspace{14mm}
   -26.1423 \eps  ^4+121.213 \eps  ^5+O(\eps ^6),
\nonumber
\displaybreak[3]\\
&\Delta_{\phi^6}=6+2 \eps -\tfrac{257}{54} \eps  ^2+17.8246 \eps  ^3
\nonumber\displaybreak[0]\\&\hspace{14mm}
-82.9519 \eps  ^4+447.314 \eps  ^5+O(\eps  ^6),
\nonumber
\displaybreak[3]\\
&\Delta_{\phi^7}=7+\tfrac72 \eps  -\tfrac{959}{108} \eps^2+38.6637 \eps^3
   \nonumber\displaybreak[0]\\&\hspace{14mm}
   -208.437 \eps^4+1291.46 \eps^5+O(\eps^6),
\nonumber
\displaybreak[3]\\
&\Delta_{(\partial \phi)^4}=8-\tfrac89 \eps  +\tfrac{22}{81} \eps^2-0.31768 \eps^3
\nonumber\displaybreak[0]\\&\hspace{14mm}
+0.749246 \eps^4
+O(\eps  ^5),
\nonumber
\displaybreak[3]\\
&\Delta_{\phi^8}=8+\tfrac{16}3 \eps  -\tfrac{1190}{81} \eps  ^2+73.4366 \eps^3
   -450.966\eps^4+O(\eps^5),
\nonumber
\\
&\Delta_{\phi(\partial \phi)^4}=9-\tfrac5{18} \eps  -\tfrac{62 }{243} \eps^2+0.812882\eps^3+O(\eps^4),
\nonumber
\\
&\Delta_{\phi^9}=9+\tfrac{15}2 \eps  -\tfrac{821}{36}\eps^2+127.189\eps^3+O(\eps^4),\nonumber
\\
&\Delta_{\partial^2(\partial \phi)^4}=10-\tfrac5{3} \eps  -\tfrac{1}{648} \eps^2+O(\eps^3),\nonumber
\\
&\Delta_{\phi^2(\partial \phi)^4}=10+\tfrac2{3} \eps  -\tfrac{269}{162} \eps^2+O(\eps^3),\nonumber
\\
&\Delta_{\phi^{10}}=10+10\eps  -\tfrac{1795}{54} \eps^2+O(\eps^3),
\end{align}
where we numerically evaluated the Riemann zeta values $\zeta_3$, $\zeta_4$, \ldots.
The five-loop result for $\phi^6$ agrees with \cite{Cao:2021cdt}, and in all other cases we improve on previous lower-order results from~\cite{Zhang1982,Derkachov:1997gc,Kehrein:1994ff,RoosmaleNepveu:2024zlz}.

\begin{figure}[t]
\includegraphics[width=0.48\textwidth]{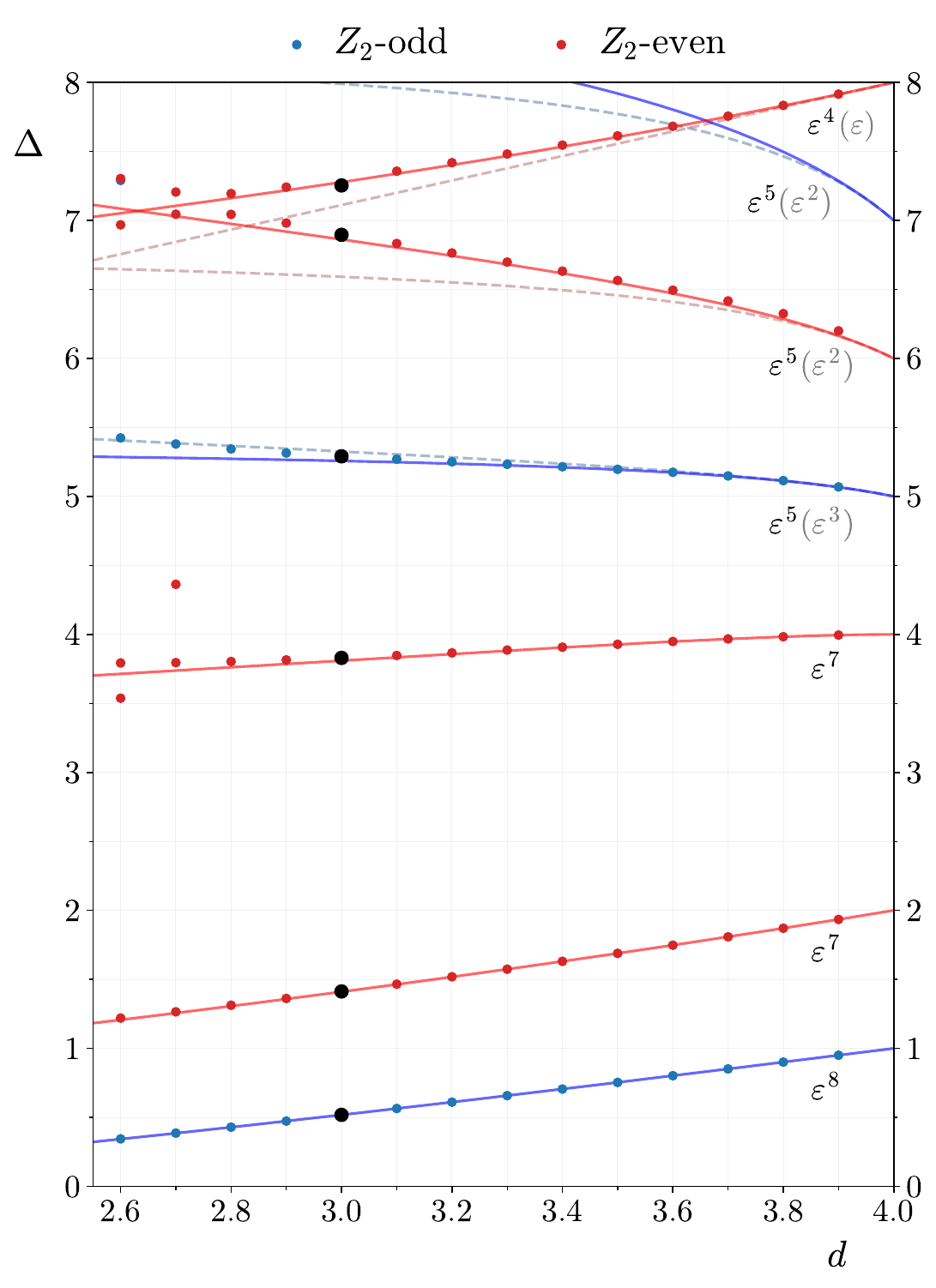}
\caption{Scalar spectrum in the Ising CFT with $\Delta<8$ for spacetime dimension $2.6\!\leq \! d\!<\!4$, using results up to $O(\eps^n)$. 
Dashed lines show previous leading determinations at $O(\textcolor{gray}{\eps^m})$. 
Numerical bootstrap results (the dots) are from \cite{Simmons-Duffin:2016wlq} (${d=3}$) and \cite{Henriksson:2022gpa}. These works did not detect the operator $\phi^7$.}
\label{fig:Ising-scalars}
\end{figure}

We present in Fig.~\ref{fig:Ising-scalars} a comprehensive picture of the scalar spectrum up to $\Delta=8$ and compare it to bootstrap results across spacetime dimensions \cite{Simmons-Duffin:2016wlq,Henriksson:2022gpa}. We find excellent agreement up to $\eps=1$ and good agreement beyond.

\subsection{\boldmath{$O(3)$} CFT and quantum-critical\\ corrections to scaling}

For the $O(n)$ CFT, we discuss the case $n=3$.
In table~\ref{tab:newO3}, we present the leading spectrum of the 3d $O(3)$ CFT, complete up to $\Delta<4$, and compare with different non-perturbative determinations directly in ${d=3}$. 

\begin{table}[t]
\caption{All operators in the 3d $O(3)$ CFT with $\Delta<4$. 
Bootstrap values are either from \cite{Chester:2020iyt} or private correspondence with the authors of that paper. 
The operator referenced with \cite{Han:2023lky} was missing in the original paper, but the authors of that paper confirm the existence of the state and gave the value 2.67 without quoting error bars. Notation: old loop order $\to$ new loop order (previous order in parenthesis). 
}
\label{tab:newO3}
\begin{tabular}{lclclll}
\hline\hline
 $R$ & $\ell$  & $\O$ & order & Pad\'e & Bootstrap & Monte Carlo 
\\\hline
$V$ & $0$ & $\phi$ & ($\eps^8$) & $.5188246$  & $.518942(51) $ & $.518920(25)$ \cite{Hasenbusch:2020pwj}
\\
$T$ & $0$ & $\phi^2$ & ($\eps^6$) & $ 1.210809$  & $1.20954(32) $ & $1.2094(3) $ \cite{Hasenbusch2011}
\\
$S$ & $0$ & $\phi^2$ & ($\eps^7$) & $1.571279 $  & $1.59489(59)$ & $1.5948(2) $ \cite{Hasenbusch:2020pwj}
\\
$A$ & $1$ & $\partial\phi^2$ & {\small{exact}} & $2 $  & 
\\
$T_3$ & $0$ & $\phi^3$ & ($\eps^6$) & $2.042931 $  & $2.03867(23)$  &  $2.0385(3) $ \cite{Hasenbusch:2022zur}
\\
$H_3$ & $1$ & $\partial\phi^3$ & $\eps\to\eps^5$ & $2.766426 $  & $2.77025(22)$ & $2.67$ \cite{Han:2023lky}  
\\
$T_4$ & $0$ & $\phi^4$ & ($\eps^6$) & $ 2.991664$  & $< 2.99056$ & $2.9857(9)$ \cite{Hasenbusch:2022zur}
\\
$S$ & $2$ & $\partial^2\phi^2$ & {\small{exact}} & $3 $  & 
\\
$T$ & $2$ & $\partial^2\phi^2$ & $\eps^4\!\to\eps^5$ & $ 3.015701$  & $3.013(18)$
\\
$V$ & $1$ & $\partial\phi^3$ & $\eps^2\!\to\eps^5$ & $3.015815 $  & $3.03120(32)$
\\
$A_3$ & $1^-$ & $\partial^2\phi^3$ & $\eps\to\eps^5$ & $3.446361 $  & NA
\\
$T$ & $0$ & $\phi^4$ & ($\eps^6$) & $3.550026 $  &  $3.561(13)$
\\
$V$ & $2$ & $\partial^2\phi^3$ & $\eps^2\!\to\eps^5$ & $ 3.630425$  & $3.633(4)$
\\
$H_4$ & $1$ & $\partial\phi^4$ & $\eps\to\eps^5$ & $3.676439 $  & $3.713(9)$
\\
$S$ & $0$ & $\phi^4$ & ($\eps^7$) & $ 3.793620$  & $3.7667(10) $ & $3.759(2)$ \cite{Hasenbusch:2020pwj}
\\
$T_3$ & $2$ & $\partial^2\phi^3$ & $\eps^2\!\to\eps^5$ & $ 3.837674$  & $3.841(19)$
\\\hline\hline
\end{tabular}
\end{table}

A particularly interesting operator in the $O(3)$ CFT is $Q=\phi^{[a}\partial_{[\mu}\phi^b\partial_{\nu]}\phi^{c]}$, which in vector notation has a component
$\vec \phi\cdot (\partial_x\vec \phi\times \partial_t\vec \phi)$. It is rank-three antisymmetric under $O(3)$ (and more generally $O(n)$) and rank-two antisymmetric under the Lorentz group, and has been proposed to contribute a substantial correction to scaling in quantum-critical behaviour described by the 3d $O(3)$ CFT \cite{Fritz2011}. We now report a five-loop result for this operator (displaying the result at $n=3$): 
\begin{align}
\Delta_Q&=5-\tfrac{3 \eps}{2}-\tfrac{15 \eps^2}{484}-\tfrac{5585 \eps^3}{234256}+\left(\tfrac{2725
   \zeta_3}{161051}-\tfrac{3507905}{340139712}\right) \eps ^4\nonumber\\&\ \ -\left(\tfrac{104650 \zeta
   _5}{1771561}-\tfrac{8175 \zeta _4}{644204}
-\tfrac{14015335 \zeta _3}{935384208}+\tfrac{1051598975}{493882861824}
\right) \eps^5\nonumber\\&\ \ +O(\eps^6),
\end{align}
where previously only the $O(\eps)$ term was known \cite{Kehrein:1994ff,Henriksson:2022rnm}. 
A Pad\'e$_{2,3}$ approximant gives $\Delta_Q^{3d}=3.44636$, of use for future studies~\cite{Parola1989,Einarsson1991,Takano2006,Wenzel2008,Ma:2018juw}.

\subsection{Cubic CFT and conformal bootstrap}

The conformal bootstrap \cite{Rattazzi:2008pe,Poland:2018epd,Rychkov:2023wsd} produces rigorous error bars for critical exponents and other conformal data, including the most precise determinations for the Ising~\cite{Chang:2024whx} and $O(n)$ \cite{Chester:2019ifh,Chester:2020iyt} CFTs.
However, results for {\em any} other scalar CFT have been considerably less precise, one reason being the increasingly complicated spectrum of less symmetric CFTs. 
Numerical bootstrap studies require input in the form of spectral gaps, which can be guided by approximate knowledge of the spectrum of the candidate theory.

Our results provide precisely this insight, which we exemplify in
$C_3 =S_3 \ltimes (Z_2)^3$ symmetric scalar field theories, referred to as cubic.
In table~\ref{tab:Cubic-ops} we present 
the entire spectrum up to engineering dimension six in the $B$ (two-index antisymmetric) and $S$ (singlet) representations of the cubic group.
These proved of crucial importance in~\cite{Kousvos:2025ext}, where a gap on the first $B$ scalar operator ($\Delta_B \geq 4.0$) was used to exclude the $O(3)$ theory from parameter space, and the leading $S$ spin-2 operator after the stress tensor ($\Delta_{T_{\mu\nu}^\prime} \geq 4.0$) was used to obtain a bootstrap island. 
Without these spectral gap assumptions, the results of \cite{Kousvos:2025ext} could not have been obtained.
Both gaps required justification by our present work, since information on these operators was not available in the literature, with the exception of the 1-loop results of \cite{Bednyakov:2023lfj}, which do not suffice for resummations.

We thus exemplified how our results enable the implementation of spectral gaps specifically in cubic theories, but we emphasise that the applicability of our theory-independent results is much broader.

\begin{table}[t]
\centering
\caption{Operators in the $B$ and $S$ representations of the cubic group. 
The used Padé resummations are: 
Pad\'e$_{3,2}$ for $B$ $\partial \phi^2$ and the first two $S$ $\partial^2 \phi^4$; 
Pad\'e$_{2,2}$ for the first $B$ $\partial \phi^4 $, 
$\partial^{[1,1]}\phi^4$, $\partial^2 \phi^4$ and the last $S$ $\partial^2 \phi^4$;
Pad\'e$_{3,3}$ for $S$ $\phi^2$ and $\phi^4$ (using \cite{Bednyakov:2023lfj}); and Padé$_{2,3}$ for all other operators.
The last column states the irrep of each operator under the $O(3)$ group.
}\label{tab:Cubic-ops}
{\small
\renewcommand{\arraystretch}{1.25}
\begin{tabular}{cllc|cllc}
\hline\hline
\multicolumn{4}{c|}{Antisymmetric $B_{[ab]}$ }&   \multicolumn{4}{c}{Singlets $S$}
\\\hline
& $\Delta_{C_3}$ & $\Delta_{O(3)}$  &${O(n)}$ & & $\Delta_{C_3}$ & $\Delta_{O(3)}$  &${O(n)}$
\\\hline
 $ \phi^6 $  &    $5.30113   $   &    $  5.27248  $   &   $ T_6 $ &  $ \phi^2 $ &    $  1.56416 $   &    $ 1.56246  $   &   $ S $
 \\ 
 $\phi^6 $    &    $5.91670  $     &    $5.93449 $  &  $  T_4$ &    $ \phi^4 $ &   $ 3.01080   $   &    $ 2.99166   $   &   $ T_4 $
 \\ \cline{1-4}
 $ \partial \phi^2$ & $2.01615$ & $ 2$ & $ A$  &     $ \phi^4 $   &    $ 3.78431   $   &    $ 3.78198   $    &  $ S $
 \\ 
 $ \partial \phi^4 $   &    $ 3.70481 $   &    $  3.68387 $    &$ H_4 $ &  $ \phi^6 $     &    $ 5.28692  $   &    $5.27248  $    &  $ T_6 $
 \\ 
 $ \partial \phi^4  $   &    $  4.09212 $   &    $ 4.09582 $  & $ A$  &  $ \square \phi^4 $   &    $ 5.02294  $ &  $ 5.02573  $    &  $ S $
 \\ \cline{1-4}
$ \! \partial^{[1,1]}\phi^4$& $ 4.45226  $&$4.44619 $&  $  A$ &  $ \phi^6 $      &    $5.89107 $   &    $ 5.93449  $   &    $ T_4 $
 \\ \cline{1-4}
 $ \partial^2 \phi^4 $    &   $4.51586 $ &    $ 4.50492  $   &    $  H_4$  &  $ \phi^6 $  &    $  6.56760 $   &    $ 6.55755 $   &     $ S $ 
 \\ \cline{5-8}
 $ \partial^2 \phi^4 $    &    $4.78803 $   &    $  4.78899$ & $T_4$  &  $ T^{\mu\nu}$ & $3$&$3$ & $S$
\\\cline{1-4} & & & & $ \partial^2 \phi^4 $     &    $ 4.73644  $   &    $4.78778  $   &  $T_4 $
 \\ & & & &   $ \partial^2 \phi^4 $ &    $ 4.77219 $   &    $  4.71329 $ &  $S $
 \\ & & & &   $ \partial^2 \phi^4 $  &    $ 5.51728  $   &    $ 5.51633  $  &  $ S $
 \\ \hline\hline
\end{tabular}
}
\end{table}

\section{Discussion and outlook}

In this paper, we have derived new multi-loop anomalous dimensions that provide a comprehensive picture of the spectrum of composite operators in scalar EFTs and CFTs. 
A central point is that, while many recent computational methods were derived with EFT applications in mind, they can also be used to derive high-quality estimates for CFT operator spectra. This then renders the study of new theories via non-perturbative methods, such as the conformal bootstrap, approachable. 
For example, bootstrapping the cubic theory of the $\varepsilon$ expansion had long remained an open problem in the bootstrap community, before perturbative intuition and our state-of-the-art data finally enabled its study~\cite{Kousvos:2025ext}. We expect that this will be the case for many other theories. Full details of our results will be presented in a subsequent publication~\cite{LongPaper}.

We used an improved $R^*$ method to access particularly high loop orders in a manageable way. 
$R^*$ is applicable to local QFTs with particles and operators of arbitrary spin and mass dimension, and for any spacetime dimension $d=n-\eps$. Furthermore, 
it reduces the complexity of integrals and it circumvents the need for explicit counterterms for subdivergences. 
The improved $R^*_{\text{ME}}$ retains all these advantages, but is more efficient, especially when applied to theories with higher-dimensional operators. The momentum expansion can also be used in a more global manner, recently to renormalize twist-2 operators in QCD \cite{Falcioni:2023luc,Falcioni:2023vqq,Falcioni:2024xyt,Falcioni:2024qpd} up to spin 20. 

Future applications of $R^*_{\text{ME}}$ within scalar theories can include targeted studies towards specific operators of interest, for instance parity-odd scalars \cite{Dymarsky:2017yzx,Zhu:2022gjc,Chang:2024whx} and singlet vector operators \cite{Meneses:2018xpu,Delamotte:2018fnz}, which are typically found at even larger engineering dimensions than considered here. Another direction would be to derive a general two-loop dilation operator, in the spirit of \cite{Kehrein:1994ff} for $O(n)$ CFTs and \cite{Beisert:2003tq,Beisert:2003jj} for $\mathcal N=4$ SYM, ideally already within the primary operator basis. 
The general results could also be used to test ideas in geometrised field spaces, e.g.~\cite{Alonso:2016oah,Helset:2022tlf}.

Having demonstrated the applicability and impact of our present results, another important extension would be to go beyond scalar theories. Relevant to this is the three-loop renormalisation of general 4d QFTs \cite{Mihaila:2013dta,Poole:2019kcm,Steudtner:2020tzo,Steudtner:2021fzs,Bednyakov:2021qxa,Davies:2021mnc,Steudtner:2024teg}, the inclusion of field potentials \cite{Jack:1982hf,Jack:1982sr,Jack:1983sk,Jack:1984vj}, and recent one-loop renormalisation of higher-dimensional operators in such theories \cite{Fonseca:2025zjb,Aebischer:2025zxg,Misiak:2025xzq}. 
Importantly, there are also CFT applications within this class, for instance 4d conformal gauge theories, including Banks--Zaks fixed-points \cite{Caswell:1974gg,Banks:1981nn,Hansen:2017pwe} and the $\varepsilon$ expansion for 3d fermionic and gauge theories \cite{Fei:2016sgs,Gracey:2016mio,Ihrig:2018hho}. 
While working with general (flavour) index structure
is the first step towards CFT applications since it allows for arbitrary global-symmetry representations, the inclusion of non-scalar operators is the next step needed to be worked out systematically (beyond twist 2).\

\begin{acknowledgments}

We thank W. Cao, G.~Guedes, A.~L\"auchli and T. Melia for useful discussions. We thank G.~Guedes, Y.\nobreakdash-T.~Huang, C.\nobreakdash-H.~Shen, A. Stergiou and A. Vichi for helpful comments on the manuscript. 
This project has received funding from the European Research Council (ERC) under the European Union's Horizon 2020 research and innovation programme (SRK under grant agreement no. 758903, JH under grant agreement number 949077). SRK also received funding from the Marie Sk\l{}odowska-Curie Action (MSCA) High energy Intelligence (HORIZON-MSCA-2023-SE-01-101182937-HeI). FH is supported by the UKRI FLF MR/Y003829/1
and the STFC Consolidated Grant ST/X000494/1. JRN is supported by the Yushan Young Scholarship 112V1039 from the Ministry of Education (MOE) of Taiwan, by the National Science and Technology Council (NSTC) grant 113-2112-M-002-038-MY3, and by the NSTC grant 114-2923-M-002-011-MY5.

\end{acknowledgments}

\bibliographystyle{ytphys}
\bibliography{bibl}

\providecommand{\href}[2]{#2}\begingroup\raggedright\begin{thebibliography}{100}

\bibitem{Grzadkowski:2010es}
B.~Grzadkowski, M.~Iskrzynski, M.~Misiak, and J.~Rosiek, {\slshape {Dimension
  six terms in the Standard Model lagrangian},}
  \href{http://dx.doi.org/10.1007/JHEP10(2010)085}{{\em JHEP} {\bfseries 10}
  (2010) 085}, \href{http://arxiv.org/abs/1008.4884}{{
  arXiv:1008.4884~[hep-ph]}}.

\bibitem{Murphy:2020rsh}
C.~W. Murphy, {\slshape {Dimension-8 operators in the Standard Model Effective
  Field Theory},} \href{http://dx.doi.org/10.1007/JHEP10(2020)174}{{\em JHEP}
  {\bfseries 10} (2020) 174}, \href{http://arxiv.org/abs/2005.00059}{{
  arXiv:2005.00059~[hep-ph]}}.

\bibitem{deVries:2019nsu}
J.~de~Vries, G.~Falcioni, F.~Herzog, and B.~Ruijl, {\slshape {Two- and
  three-loop anomalous dimensions of Weinberg{\textquoteright}s dimension-six
  CP-odd gluonic operator},}
  \href{http://dx.doi.org/10.1103/PhysRevD.102.016010}{{\em Phys. Rev. D}
  {\bfseries 102} (2020) 016010}, \href{http://arxiv.org/abs/1907.04923}{{
  arXiv:1907.04923~[hep-ph]}}.

\bibitem{Aebischer:2022anv}
J.~Aebischer, A.~J. Buras, and J.~Kumar, {\slshape {NLO QCD renormalization
  group evolution for nonleptonic {\ensuremath{\Delta}}F=2 transitions in the
  SMEFT},} \href{http://dx.doi.org/10.1103/PhysRevD.106.035003}{{\em Phys. Rev.
  D} {\bfseries 106} (2022) 035003}, \href{http://arxiv.org/abs/2203.11224}{{
  arXiv:2203.11224~[hep-ph]}}.

\bibitem{Jenkins:2023bls}
E.~E. Jenkins, A.~V. Manohar, L.~Naterop, and J.~Pag\`es, {\slshape {Two loop
  renormalization of scalar theories using a geometric approach},}
  \href{http://dx.doi.org/10.1007/JHEP02(2024)131}{{\em JHEP} {\bfseries 02}
  (2024) 131}, \href{http://arxiv.org/abs/2310.19883}{{
  arXiv:2310.19883~[hep-ph]}}.

\bibitem{Ibarra:2024tpt}
A.~Ibarra, N.~Leister, and D.~Zhang, {\slshape {Complete two-loop
  renormalization group equation of the Weinberg operator},}
  \href{http://dx.doi.org/10.1007/JHEP03(2025)214}{{\em JHEP} {\bfseries 03}
  (2025) 214}, \href{http://arxiv.org/abs/2411.08011}{{
  arXiv:2411.08011~[hep-ph]}}.

\bibitem{DiNoi:2024ajj}
S.~Di~Noi, R.~Gr\"ober, and M.~K. Mandal, {\slshape {Two-loop running effects
  in Higgs physics in Standard Model Effective Field Theory},}
  \href{http://dx.doi.org/10.1007/JHEP12(2024)220}{{\em JHEP} {\bfseries 12}
  (2025) 220}, \href{http://arxiv.org/abs/2408.03252}{{
  arXiv:2408.03252~[hep-ph]}}.

\bibitem{Born:2024mgz}
L.~Born, J.~Fuentes-Mart\'\i{}n, S.~Kvedarait\.{e}, and A.~E. Thomsen,
  {\slshape {Two-loop running in the bosonic SMEFT using functional methods},}
  \href{http://dx.doi.org/10.1007/JHEP05(2025)121}{{\em JHEP} {\bfseries 05}
  (2025) 121}, \href{http://arxiv.org/abs/2410.07320}{{
  arXiv:2410.07320~[hep-ph]}}.

\bibitem{Duhr:2025zqw}
C.~Duhr, A.~Vasquez, G.~Ventura, and E.~Vryonidou, {\slshape {Two-loop
  renormalisation of quark and gluon fields in the SMEFT},}
  \href{http://arxiv.org/abs/2503.01954}{{ arXiv:2503.01954~[hep-ph]}}.

\bibitem{Zhang:2025ywe}
D.~Zhang, {\slshape {Two-loop renormalization group equations in the
  $\nu$SMEFT},} \href{http://arxiv.org/abs/2504.00792}{{
  arXiv:2504.00792~[hep-ph]}}.

\bibitem{Ferrara:1973yt}
S.~Ferrara, A.~F. Grillo, and R.~Gatto, {\slshape {Tensor representations of
  conformal algebra and conformally covariant operator product expansion},}
  \href{http://dx.doi.org/10.1016/0003-4916(73)90446-6}{{\em Annals Phys.}
  {\bfseries 76} (1973) 161--188}.

\bibitem{Osborn:1993cr}
H.~Osborn and A.~C. Petkou, {\slshape {Implications of conformal invariance in
  field theories for general dimensions},}
  \href{http://dx.doi.org/10.1006/aphy.1994.1045}{{\em Annals Phys.} {\bfseries
  231} (1994) 311--362}, \href{http://arxiv.org/abs/hep-th/9307010}{{
  arXiv:hep-th/9307010}}.

\bibitem{Rychkov:2016iqz}
S.~Rychkov, {\slshape {EPFL lectures on conformal field theory in $D\geq 3$
  dimensions},} \href{http://dx.doi.org/10.1007/978-3-319-43626-5}{{\em
  SpringerBriefs Phys.} (2017) }, \href{http://arxiv.org/abs/1601.05000}{{
  arXiv:1601.05000~[hep-th]}}.

\bibitem{Simmons-Duffin:2016gjk}
D.~Simmons-Duffin,
  \href{http://dx.doi.org/10.1142/9789813149441\_0001}{{\slshape {TASI lectures
  on the conformal bootstrap},}} in {\em {Theoretical Advanced Study Institute
  in Elementary Particle Physics: New Frontiers in Fields and Strings}}.
\newblock 2015.
\newblock \href{http://arxiv.org/abs/1602.07982}{{ arXiv:1602.07982~[hep-th]}}.

\bibitem{Pelissetto:2000ek}
A.~Pelissetto and E.~Vicari, {\slshape {Critical phenomena and renormalization
  group theory},} \href{http://dx.doi.org/10.1016/S0370-1573(02)00219-3}{{\em
  Phys. Rept.} {\bfseries 368} (2002) 549--727},
  \href{http://arxiv.org/abs/cond-mat/0012164}{{ arXiv:cond-mat/0012164}}.

\bibitem{Wilson:1971dc}
K.~G. Wilson and M.~E. Fisher, {\slshape {Critical exponents in $3.99$
  dimensions},} \href{http://dx.doi.org/10.1103/PhysRevLett.28.240}{{\em Phys.
  Rev. Lett.} {\bfseries 28} (1972) 240--243}.

\bibitem{Wilson:1973jj}
K.~G. Wilson and J.~B. Kogut, {\slshape {The Renormalization group and the
  $\epsilon$ expansion},}
  \href{http://dx.doi.org/10.1016/0370-1573(74)90023-4}{{\em Phys. Rept.}
  {\bfseries 12} (1974) 75--199}.

\bibitem{Bednyakov:2021ojn}
A.~Bednyakov and A.~Pikelner, {\slshape {Six-loop beta functions in general
  scalar theory},} \href{http://dx.doi.org/10.1007/JHEP04(2021)233}{{\em JHEP}
  {\bfseries 04} (2021) 233}, \href{http://arxiv.org/abs/2102.12832}{{
  arXiv:2102.12832~[hep-ph]}}.

\bibitem{Schnetz:2022nsc}
O.~Schnetz, {\slshape {$\phi^4$ theory at seven loops},}
  \href{http://dx.doi.org/10.1103/PhysRevD.107.036002}{{\em Phys. Rev. D}
  {\bfseries 107} (2023) 036002}, \href{http://arxiv.org/abs/2212.03663}{{
  arXiv:2212.03663~[hep-th]}}.

\bibitem{Henning:2015alf}
B.~Henning, X.~Lu, T.~Melia, and H.~Murayama, {\slshape {$2$, $84$, $30$,
  $993$, $560$, $15456$, $11962$, $261485$, \ldots: Higher dimension operators
  in the SM EFT},} \href{http://dx.doi.org/10.1007/JHEP08(2017)016}{{\em JHEP}
  {\bfseries 08} (2017) 016}, \href{http://arxiv.org/abs/1512.03433}{{
  arXiv:1512.03433~[hep-ph]}}. [Erratum: {\em JHEP } {\bf 09} (2019) 019].

\bibitem{Henning:2017fpj}
B.~Henning, X.~Lu, T.~Melia, and H.~Murayama, {\slshape {Operator bases,
  $S$-matrices, and their partition functions},}
  \href{http://dx.doi.org/10.1007/JHEP10(2017)199}{{\em JHEP} {\bfseries 10}
  (2017) 199}, \href{http://arxiv.org/abs/1706.08520}{{
  arXiv:1706.08520~[hep-th]}}.

\bibitem{Cao:2021cdt}
W.~Cao, F.~Herzog, T.~Melia, and J.~R. Nepveu, {\slshape {Renormalization and
  non-renormalization of scalar EFTs at higher orders},}
  \href{http://dx.doi.org/10.1007/JHEP09(2021)014}{{\em JHEP} {\bfseries 09}
  (2021) 014}, \href{http://arxiv.org/abs/2105.12742}{{
  arXiv:2105.12742~[hep-ph]}}.

\bibitem{Vladimirov:1979zm}
A.~A. Vladimirov, {\slshape {Method for computing renormalization group
  functions in dimensional renormalization scheme},}
  \href{http://dx.doi.org/10.1007/BF01018394}{{\em Theor. Math. Phys.}
  {\bfseries 43} (1980) 417}.

\bibitem{Chetyrkin:1982nn}
K.~G. Chetyrkin and F.~V. Tkachov, {\slshape {Infrared R operation and
  ultrviolet counterterms in the MS scheme},}
  \href{http://dx.doi.org/10.1016/0370-2693(82)90358-6}{{\em Phys. Lett. B}
  {\bfseries 114} (1982) 340--344}.

\bibitem{Chetyrkin:1984xa}
K.~G. Chetyrkin and V.~A. Smirnov, {\slshape {R* operation corrected},}
  \href{http://dx.doi.org/10.1016/0370-2693(84)91291-7}{{\em Phys. Lett. B}
  {\bfseries 144} (1984) 419--424}.

\bibitem{Smirnov:1985yck}
V.~A. Smirnov and K.~G. Chetyrkin, {\slshape {$R^*$ operation in the Minimal
  Subtraction Scheme},} \href{http://dx.doi.org/10.1007/BF01017902}{{\em Theor.
  Math. Phys.} {\bfseries 63} (1985) 462--469}.

\bibitem{Larin:2002sc}
S.~Larin and P.~van Nieuwenhuizen, {\slshape {The Infrared R* operation},}
  \href{http://arxiv.org/abs/hep-th/0212315}{{ arXiv:hep-th/0212315}}.

\bibitem{Kleinert:2001hn}
H.~Kleinert and V.~Schulte-Frohlinde,
  \href{http://dx.doi.org/10.1142/4733}{{\em {Critical Properties of
  $\phi^4$-Theories}}}.
\newblock World Scientific Publishing, 2001.

\bibitem{Chetyrkin:2017ppe}
K.~G. Chetyrkin, {\slshape {Combinatorics of $\mathbf{R}$-, $\mathbf{R^{-1}}$-,
  and $\mathbf{R^*}$-operations and asymptotic expansions of feynman integrals
  in the limit of large momenta and masses},}
  \href{http://arxiv.org/abs/1701.08627}{{ arXiv:1701.08627~[hep-th]}}.

\bibitem{Herzog:2017bjx}
F.~Herzog and B.~Ruijl, {\slshape {The R$^{*}$-operation for Feynman graphs
  with generic numerators},}
  \href{http://dx.doi.org/10.1007/JHEP05(2017)037}{{\em JHEP} {\bfseries 05}
  (2017) 037}, \href{http://arxiv.org/abs/1703.03776}{{
  arXiv:1703.03776~[hep-th]}}.

\bibitem{Beekveldt:2020kzk}
R.~Beekveldt, M.~Borinsky, and F.~Herzog, {\slshape {The Hopf algebra structure
  of the R$^{*}$-operation},}
  \href{http://dx.doi.org/10.1007/JHEP07(2020)061}{{\em JHEP} {\bfseries 07}
  (2020) 061}, \href{http://arxiv.org/abs/2003.04301}{{
  arXiv:2003.04301~[hep-th]}}.

\bibitem{Chakraborty2023}
M.~Chakraborty, {\slshape {The Asymptotic Hopf Algebra of Feynman Integrals},}
  master's thesis, Indian Institute of Science, 2023.

\bibitem{Chetyrkin:1982zq}
K.~G. Chetyrkin, F.~V. Tkachov, and S.~G. Gorishnii, {\slshape {Operator
  product expansion in the minimal subtraction scheme},}
  \href{http://dx.doi.org/10.1016/0370-2693(82)90701-8}{{\em Phys. Lett. B}
  {\bfseries 119} (1982) 407--411}.

\bibitem{Chetyrkin:1983qlc}
K.~G. Chetyrkin, {\slshape {Infrared R\ensuremath{*} - operation and operator
  product expansion in the minimal subtraction scheme},}
  \href{http://dx.doi.org/10.1016/0370-2693(83)90183-1}{{\em Phys. Lett. B}
  {\bfseries 126} (1983) 371--375}.

\bibitem{Gorishnii:1983su}
S.~G. Gorishnii, S.~A. Larin, and F.~V. Tkachov, {\slshape {The algorithm for
  OPE coefficient functions in the MS scheme},}
  \href{http://dx.doi.org/10.1016/0370-2693(83)91439-9}{{\em Phys. Lett. B}
  {\bfseries 124} (1983) 217--220}.

\bibitem{Gorishnii:1986mv}
S.~G. Gorishnii, {\slshape {On the construction of operator expansions and
  effective theories in the MS scheme. Examples: Infrared finiteness of
  coefficient functions},} {\em Dubna preprint} (1986) [JINR--E2--86--177].

\bibitem{LlewellynSmith:1987jx}
C.~H. Llewellyn~Smith and J.~P. de~Vries, {\slshape {The operator product
  expansion for minimally subtracted operators},}
  \href{http://dx.doi.org/10.1016/0550-3213(88)90407-5}{{\em Nucl. Phys. B}
  {\bfseries 296} (1988) 991--1006}.

\bibitem{Gorishnii:1986gn}
S.~G. Gorishnii and S.~A. Larin, {\slshape {Coefficient functions of asymptotic
  operator expansions in minimal subtraction scheme},}
  \href{http://dx.doi.org/10.1016/0550-3213(87)90283-5}{{\em Nucl. Phys. B}
  {\bfseries 283} (1987) 452}.

\bibitem{Chetyrkin:1988zz}
K.~G. Chetyrkin, {\slshape {Operator expansions in the minimal subtraction
  scheme. 1: The gluing method},}
  \href{http://dx.doi.org/10.1007/BF01017168}{{\em Theor. Math. Phys.}
  {\bfseries 75} (1988) 346--356}.

\bibitem{Chetyrkin:1988cu}
K.~G. Chetyrkin, {\slshape {Operator expansions in the minimal subtraction
  scheme. 2: Explicit formulas for coefficient functions},}
  \href{http://dx.doi.org/10.1007/BF01028580}{{\em Theor. Math. Phys.}
  {\bfseries 76} (1988) 809--817}.

\bibitem{Gorishnii:1989dd}
S.~G. Gorishnii, {\slshape {Construction of operator expansions and effective
  theories in the MS scheme},}
  \href{http://dx.doi.org/10.1016/0550-3213(89)90622-6}{{\em Nucl. Phys. B}
  {\bfseries 319} (1989) 633--666}.

\bibitem{Smirnov:1990rz}
V.~A. Smirnov, {\slshape {Asymptotic expansions in limits of large momenta and
  masses},} \href{http://dx.doi.org/10.1007/BF02102092}{{\em Commun. Math.
  Phys.} {\bfseries 134} (1990) 109--137}.

\bibitem{Smirnov:1994tg}
V.~A. Smirnov, {\slshape {Asymptotic expansions in momenta and masses and
  calculation of Feynman diagrams},}
  \href{http://dx.doi.org/10.1142/S0217732395001617}{{\em Mod. Phys. Lett. A}
  {\bfseries 10} (1995) 1485--1500},
  \href{http://arxiv.org/abs/hep-th/9412063}{{ arXiv:hep-th/9412063}}.

\bibitem{Smirnov:2002pj}
V.~A. Smirnov, {\slshape {Applied asymptotic expansions in momenta and
  masses},} \href{http://dx.doi.org/10.1007/3-540-44574-9}{{\em Springer Tracts
  Mod. Phys.} {\bfseries 177} (2002) 1--262}.

\bibitem{Chakraborty:2024uzz}
M.~Chakraborty and F.~Herzog, {\slshape {The asymptotic Hopf algebra of Feynman
  integrals},} \href{http://dx.doi.org/10.1007/JHEP01(2025)006}{{\em JHEP}
  {\bfseries 01} (2025) 006}, \href{http://arxiv.org/abs/2408.14304}{{
  arXiv:2408.14304~[hep-th]}}.

\bibitem{Bogoliubov:1957gp}
N.~N. Bogoliubov and O.~S. Parasiuk, {\slshape {On the Multiplication of the
  causal function in the quantum theory of fields},}
  \href{http://dx.doi.org/10.1007/BF02392399}{{\em Acta Math.} {\bfseries 97}
  (1957) 227--266}.

\bibitem{Hepp:1966eg}
K.~Hepp, {\slshape {Proof of the Bogolyubov-Parasiuk theorem on
  renormalization},} \href{http://dx.doi.org/10.1007/BF01773358}{{\em Commun.
  Math. Phys.} {\bfseries 2} (1966) 301--326}.

\bibitem{Zimmermann:1969jj}
W.~Zimmermann, {\slshape {Convergence of Bogolyubov's method of renormalization
  in momentum space},} \href{http://dx.doi.org/10.1007/BF01645676}{{\em Commun.
  Math. Phys.} {\bfseries 15} (1969) 208--234}.

\bibitem{Caswell:1981ek}
W.~E. Caswell and A.~D. Kennedy, {\slshape {A simple approach to
  renormalization theory},}
  \href{http://dx.doi.org/10.1103/PhysRevD.25.392}{{\em Phys. Rev. D}
  {\bfseries 25} (1982) 392}.

\bibitem{maple}
{Maplesoft, a division of Waterloo Maple Inc..}, ``Maple.''
\newblock \url{https://hadoop.apache.org}.

\bibitem{Vermaseren:2000nd}
J.~A.~M. Vermaseren, {\slshape {New features of FORM},}
  \href{http://arxiv.org/abs/math-ph/0010025}{{ arXiv:math-ph/0010025}}.

\bibitem{Ruijl:2017dtg}
B.~Ruijl, T.~Ueda, and J.~Vermaseren, {\slshape {FORM version 4.2},}
  \href{http://arxiv.org/abs/1707.06453}{{ arXiv:1707.06453~[hep-ph]}}.

\bibitem{Baikov:2010hf}
P.~A. Baikov and K.~G. Chetyrkin, {\slshape {Four loop massless propagators: An
  algebraic evaluation of all master integrals},}
  \href{http://dx.doi.org/10.1016/j.nuclphysb.2010.05.004}{{\em Nucl. Phys. B}
  {\bfseries 837} (2010) 186--220}, \href{http://arxiv.org/abs/1004.1153}{{
  arXiv:1004.1153~[hep-ph]}}.

\bibitem{Lee:2011jt}
R.~N. Lee, A.~V. Smirnov, and V.~A. Smirnov, {\slshape {Master Integrals for
  Four-Loop Massless Propagators up to Transcendentality Weight Twelve},}
  \href{http://dx.doi.org/10.1016/j.nuclphysb.2011.11.005}{{\em Nucl. Phys. B}
  {\bfseries 856} (2012) 95--110}, \href{http://arxiv.org/abs/1108.0732}{{
  arXiv:1108.0732~[hep-th]}}.

\bibitem{Ruijl:2017cxj}
B.~Ruijl, T.~Ueda, and J.~A.~M. Vermaseren, {\slshape {Forcer, a FORM program
  for the parametric reduction of four-loop massless propagator diagrams},}
  \href{http://dx.doi.org/10.1016/j.cpc.2020.107198}{{\em Comput. Phys.
  Commun.} {\bfseries 253} (2020) 107198},
  \href{http://arxiv.org/abs/1704.06650}{{ arXiv:1704.06650~[hep-ph]}}.

\bibitem{Goode:2024cfy}
J.~Goode, F.~Herzog, and S.~Teale, {\slshape {OPITeR: A program for tensor
  reduction of multi-loop Feynman integrals},}
  \href{http://dx.doi.org/10.1016/j.cpc.2025.109606}{{\em Comput. Phys.
  Commun.} {\bfseries 312} (2025) 109606},
  \href{http://arxiv.org/abs/2411.02233}{{ arXiv:2411.02233~[hep-ph]}}.

\bibitem{Kaneko:1994fd}
T.~Kaneko, {\slshape {A Feynman graph generator for any order of coupling
  constants},} \href{http://dx.doi.org/10.1016/0010-4655(95)00122-6}{{\em
  Comput. Phys. Commun.} {\bfseries 92} (1995) 127--152},
  \href{http://arxiv.org/abs/hep-th/9408107}{{ arXiv:hep-th/9408107}}.

\bibitem{Kaneko:2017wzd}
T.~Kaneko, {\slshape {Counting the number of Feynman graphs in QCD},}
  \href{http://dx.doi.org/10.1016/j.cpc.2017.12.020}{{\em Comput. Phys.
  Commun.} {\bfseries 226} (2018) 104--113},
  \href{http://arxiv.org/abs/1709.02067}{{ arXiv:1709.02067~[hep-ph]}}.

\bibitem{Henriksson:2022rnm}
J.~Henriksson, {\slshape {The critical $\mathrm O(N)$ CFT: Methods and
  conformal data},} \href{http://dx.doi.org/10.1016/j.physrep.2022.12.002}{{\em
  Phys. Rept.} {\bfseries 1002} (2023) 1--72},
  \href{http://arxiv.org/abs/2201.09520}{{ arXiv:2201.09520~[hep-th]}}.

\bibitem{Bednyakov:2023lfj}
A.~Bednyakov, J.~Henriksson, and S.~R. Kousvos, {\slshape {Anomalous dimensions
  in hypercubic theories},}
  \href{http://dx.doi.org/10.1007/JHEP11(2023)051}{{\em JHEP} {\bfseries 11}
  (2023) 051}, \href{http://arxiv.org/abs/2304.06755}{{
  arXiv:2304.06755~[hep-th]}}.

\bibitem{Derkachov:1997gc}
S.~E. Derkachov and A.~N. Manashov, {\slshape {On the stability problem in the
  $O(N)$ nonlinear sigma model},}
  \href{http://dx.doi.org/10.1103/PhysRevLett.79.1423}{{\em Phys. Rev. Lett.}
  {\bfseries 79} (1997) 1423--1427},
  \href{http://arxiv.org/abs/hep-th/9705020}{{ arXiv:hep-th/9705020}}.

\bibitem{RoosmaleNepveu:2024zlz}
J.~Roosmale~Nepveu, \href{http://dx.doi.org/10.18452/29107}{{\em
  {Renormalization and the Double Copy of Effective Field Theories}}}.
\newblock PhD thesis, Humboldt U., Berlin, 2024.

\bibitem{LongPaper}
J.~Henriksson, S.~R. Kousvos, and J.~Roosmale~Nepveu, {\slshape {EFT meets CFT:
  Multi-loop renormalization of higher-dimensional operators in general
  $\phi^4$ theories},} {\em (Work in progress)} (2025).

\bibitem{Zhang1982}
F.~C. Zhang and R.~K.~P. Zia, {\slshape {A correction-to-scaling critical
  exponent for fluids at order $\epsilon^3$},}
  \href{http://dx.doi.org/10.1088/0305-4470/15/10/032}{{\em J. Phys. A}
  {\bfseries 15} (1982) 3303--3305}.

\bibitem{Bertucci:2022ptt}
F.~Bertucci, J.~Henriksson, and B.~McPeak, {\slshape {Analytic bootstrap of
  mixed correlators in the $O(n)$ CFT},}
  \href{http://dx.doi.org/10.1007/JHEP10(2022)104}{{\em JHEP} {\bfseries 10}
  (2022) 104}, \href{http://arxiv.org/abs/2205.09132}{{
  arXiv:2205.09132~[hep-th]}}.

\bibitem{Kehrein:1994ff}
S.~K. Kehrein and F.~Wegner, {\slshape {The structure of the spectrum of
  anomalous dimensions in the $N$-vector model in $4-\epsilon$ dimensions},}
  \href{http://dx.doi.org/10.1016/0550-3213(94)90406-5}{{\em Nucl. Phys. B}
  {\bfseries 424} (1994) 521--546},
  \href{http://arxiv.org/abs/hep-th/9405123}{{ arXiv:hep-th/9405123}}.

\bibitem{Simmons-Duffin:2016wlq}
D.~Simmons-Duffin, {\slshape {The lightcone bootstrap and the spectrum of the
  $3$d Ising CFT},} \href{http://dx.doi.org/10.1007/JHEP03(2017)086}{{\em JHEP}
  {\bfseries 03} (2017) 086}, \href{http://arxiv.org/abs/1612.08471}{{
  arXiv:1612.08471~[hep-th]}}.

\bibitem{Reehorst:2021hmp}
M.~Reehorst, {\slshape {Rigorous bounds on irrelevant operators in the $3$d
  Ising model CFT},} \href{http://dx.doi.org/10.1007/JHEP09(2022)177}{{\em
  JHEP} {\bfseries 09} (2022) 177}, \href{http://arxiv.org/abs/2111.12093}{{
  arXiv:2111.12093~[hep-th]}}.

\bibitem{Henriksson:2022gpa}
J.~Henriksson, S.~R. Kousvos, and M.~Reehorst, {\slshape {Spectrum continuity
  and level repulsion: the Ising CFT from infinitesimal to finite
  $\varepsilon$},} \href{http://dx.doi.org/10.1007/JHEP02(2023)218}{{\em JHEP}
  {\bfseries 02} (2023) 218}, \href{http://arxiv.org/abs/2207.10118}{{
  arXiv:2207.10118~[hep-th]}}.

\bibitem{Chester:2020iyt}
S.~M. Chester, W.~Landry, J.~Liu, D.~Poland, D.~Simmons-Duffin, N.~Su, and
  A.~Vichi, {\slshape {Bootstrapping Heisenberg magnets and their cubic
  instability},} \href{http://dx.doi.org/10.1103/PhysRevD.104.105013}{{\em
  Phys. Rev. D} {\bfseries 104} (2021) 105013},
  \href{http://arxiv.org/abs/2011.14647}{{ arXiv:2011.14647~[hep-th]}}.

\bibitem{Han:2023lky}
C.~Han, L.~Hu, and W.~Zhu, {\slshape {Conformal operator content of the
  Wilson-Fisher transition on fuzzy sphere bilayers},}
  \href{http://dx.doi.org/10.1103/PhysRevB.110.115113}{{\em Phys. Rev. B}
  {\bfseries 110} (2024) 115113}, \href{http://arxiv.org/abs/2312.04047}{{
  arXiv:2312.04047~[cond-mat.str-el]}}.

\bibitem{Hasenbusch:2020pwj}
M.~Hasenbusch, {\slshape {Monte Carlo study of a generalized icosahedral model
  on the simple cubic lattice},}
  \href{http://dx.doi.org/10.1103/PhysRevB.102.024406}{{\em Phys. Rev. B}
  {\bfseries 102} (2020) 024406}, \href{http://arxiv.org/abs/2005.04448}{{
  arXiv:2005.04448~[cond-mat.stat-mech]}}.

\bibitem{Hasenbusch2011}
M.~Hasenbusch and E.~Vicari, {\slshape {Anisotropic perturbations in
  three-dimensional $\mathrm O(N)$-symmetric vector models},}
  \href{http://dx.doi.org/10.1103/physrevb.84.125136}{{\em Phys. Rev. B}
  {\bfseries 84} (2011) 125136}, \href{http://arxiv.org/abs/1108.0491}{{
  arXiv:1108.0491}}.

\bibitem{Hasenbusch:2022zur}
M.~Hasenbusch, {\slshape {Cubic fixed point in three dimensions: Monte Carlo
  simulations of the $\phi^4$ model on the simple cubic lattice},}
  \href{http://dx.doi.org/10.1103/PhysRevB.107.024409}{{\em Phys. Rev. B}
  {\bfseries 107} (2023) 024409}, \href{http://arxiv.org/abs/2211.16170}{{
  arXiv:2211.16170~[cond-mat.stat-mech]}}.

\bibitem{Fritz2011}
L.~Fritz, R.~L. Doretto, S.~Wessel, S.~Wenzel, S.~Burdin, and M.~Vojta,
  {\slshape {Anomalous quantum-critical scaling corrections in two-dimensional
  antiferromagnets},} \href{http://dx.doi.org/10.1103/PhysRevB.83.174416}{{\em
  Phys. Rev. B} {\bfseries 83} (2011) 174416},
  \href{http://arxiv.org/abs/1101.3784}{{ arXiv:1101.3784~[cond-mat.str-el]}}.

\bibitem{Parola1989}
A.~Parola, {\slshape {Effective Hamiltonians for Heisenberg antiferromagnets},}
  \href{http://dx.doi.org/10.1103/PhysRevB.40.7109}{{\em Phys. Rev. B}
  {\bfseries 40} (1989) 7109--7114}.

\bibitem{Einarsson1991}
T.~Einarsson and H.~Johannesson, {\slshape {Effective-action approach to the
  frustrated Heisenberg antiferromagnet in two dimensions},}
  \href{http://dx.doi.org/10.1103/PhysRevB.43.5867}{{\em Phys. Rev. B}
  {\bfseries 43} (1991) 5867--5882}.

\bibitem{Takano2006}
K.~Takano, {\slshape {Spin-gap phase of a quantum spin system on a honeycomb
  lattice},} \href{http://dx.doi.org/10.1103/PhysRevB.74.140402}{{\em Phys.
  Rev. B} {\bfseries 74} (2006) 140402},
  \href{http://arxiv.org/abs/cond-mat/0609446}{{
  arXiv:cond-mat/0609446~[cond-mat.str-el]}}.

\bibitem{Wenzel2008}
S.~Wenzel, L.~Bogacz, and W.~Janke, {\slshape {Evidence for an unconventional
  universality class from a two-dimensional dimerized quantum Heisenberg
  model},} \href{http://dx.doi.org/10.1103/PhysRevLett.101.127202}{{\em Phys.
  Rev. Lett.} {\bfseries 101} (2008) 127202},
  \href{http://arxiv.org/abs/0805.2500}{{
  arXiv:0805.2500~[cond-mat.stat-mech]}}.

\bibitem{Ma:2018juw}
N.~Ma, P.~Weinberg, H.~Shao, W.~Guo, D.-X. Yao, and A.~W. Sandvik, {\slshape
  {Anomalous quantum-critical scaling corrections in two-dimensional
  antiferromagnets},}
  \href{http://dx.doi.org/10.1103/PhysRevLett.121.117202}{{\em Phys. Rev.
  Lett.} {\bfseries 121} (2018) 117202},
  \href{http://arxiv.org/abs/1804.01273}{{
  arXiv:1804.01273~[cond-mat.str-el]}}.

\bibitem{Rattazzi:2008pe}
R.~Rattazzi, V.~S. Rychkov, E.~Tonni, and A.~Vichi, {\slshape {Bounding scalar
  operator dimensions in 4D CFT},}
  \href{http://dx.doi.org/10.1088/1126-6708/2008/12/031}{{\em JHEP} {\bfseries
  12} (2008) 031}, \href{http://arxiv.org/abs/0807.0004}{{
  arXiv:0807.0004~[hep-th]}}.

\bibitem{Poland:2018epd}
D.~Poland, S.~Rychkov, and A.~Vichi, {\slshape {The Conformal Bootstrap:
  Theory, numerical techniques, and applications},}
  \href{http://dx.doi.org/10.1103/RevModPhys.91.015002}{{\em Rev. Mod. Phys.}
  {\bfseries 91} (2019) 015002}, \href{http://arxiv.org/abs/1805.04405}{{
  arXiv:1805.04405~[hep-th]}}.

\bibitem{Rychkov:2023wsd}
S.~Rychkov and N.~Su, {\slshape {New developments in the numerical conformal
  bootstrap},} \href{http://dx.doi.org/10.1103/RevModPhys.96.045004}{{\em Rev.
  Mod. Phys.} {\bfseries 96} (2024) 045004},
  \href{http://arxiv.org/abs/2311.15844}{{ arXiv:2311.15844~[hep-th]}}.

\bibitem{Chang:2024whx}
C.-H. Chang, V.~Dommes, R.~S. Erramilli, A.~Homrich, P.~Kravchuk, A.~Liu, M.~S.
  Mitchell, D.~Poland, and D.~Simmons-Duffin, {\slshape {Bootstrapping the 3d
  Ising stress tensor},} \href{http://dx.doi.org/10.1007/JHEP03(2025)136}{{\em
  JHEP} {\bfseries 03} (2025) 136}, \href{http://arxiv.org/abs/2411.15300}{{
  arXiv:2411.15300~[hep-th]}}.

\bibitem{Chester:2019ifh}
S.~M. Chester, W.~Landry, J.~Liu, D.~Poland, D.~Simmons-Duffin, N.~Su, and
  A.~Vichi, {\slshape {Carving out OPE space and precise $O(2)$ model critical
  exponents},} \href{http://dx.doi.org/10.1007/JHEP06(2020)142}{{\em JHEP}
  {\bfseries 06} (2020) 142}, \href{http://arxiv.org/abs/1912.03324}{{
  arXiv:1912.03324~[hep-th]}}.

\bibitem{Kousvos:2025ext}
S.~R. Kousvos and A.~Stergiou, {\slshape {Redundancy channels in the conformal
  bootstrap},} \href{http://arxiv.org/abs/2507.05338}{{
  arXiv:2507.05338~[hep-th]}}.

\bibitem{Falcioni:2023luc}
G.~Falcioni, F.~Herzog, S.~Moch, and A.~Vogt, {\slshape {Four-loop splitting
  functions in QCD {\textendash} The quark-quark case},}
  \href{http://dx.doi.org/10.1016/j.physletb.2023.137944}{{\em Phys. Lett. B}
  {\bfseries 842} (2023) 137944}, \href{http://arxiv.org/abs/2302.07593}{{
  arXiv:2302.07593~[hep-ph]}}.

\bibitem{Falcioni:2023vqq}
G.~Falcioni, F.~Herzog, S.~Moch, and A.~Vogt, {\slshape {Four-loop splitting
  functions in QCD {\textendash} The gluon-to-quark case},}
  \href{http://dx.doi.org/10.1016/j.physletb.2023.138215}{{\em Phys. Lett. B}
  {\bfseries 846} (2023) 138215}, \href{http://arxiv.org/abs/2307.04158}{{
  arXiv:2307.04158~[hep-ph]}}.

\bibitem{Falcioni:2024xyt}
G.~Falcioni, F.~Herzog, S.~Moch, A.~Pelloni, and A.~Vogt, {\slshape {Four-loop
  splitting functions in QCD: The quark-to-gluon case},}
  \href{http://dx.doi.org/10.1016/j.physletb.2024.138906}{{\em Phys. Lett. B}
  {\bfseries 856} (2024) 138906}, \href{http://arxiv.org/abs/2404.09701}{{
  arXiv:2404.09701~[hep-ph]}}.

\bibitem{Falcioni:2024qpd}
G.~Falcioni, F.~Herzog, S.~Moch, A.~Pelloni, and A.~Vogt, {\slshape {Four-loop
  splitting functions in QCD: the gluon-gluon case},}
  \href{http://dx.doi.org/10.1016/j.physletb.2024.139194}{{\em Phys. Lett. B}
  {\bfseries 860} (2025) 139194}, \href{http://arxiv.org/abs/2410.08089}{{
  arXiv:2410.08089~[hep-ph]}}.

\bibitem{Dymarsky:2017yzx}
A.~Dymarsky, F.~Kos, P.~Kravchuk, D.~Poland, and D.~Simmons-Duffin, {\slshape
  {The $3$d stress-tensor bootstrap},}
  \href{http://dx.doi.org/10.1007/JHEP02(2018)164}{{\em JHEP} {\bfseries 02}
  (2018) 164}, \href{http://arxiv.org/abs/1708.05718}{{
  arXiv:1708.05718~[hep-th]}}.

\bibitem{Zhu:2022gjc}
W.~Zhu, C.~Han, E.~Huffman, J.~S. Hofmann, and Y.-C. He, {\slshape {Uncovering
  conformal symmetry in the 3D Ising transition: State-operator correspondence
  from a quantum fuzzy sphere regularization},}
  \href{http://dx.doi.org/10.1103/PhysRevX.13.021009}{{\em Phys. Rev. X}
  {\bfseries 13} (2023) 021009}, \href{http://arxiv.org/abs/2210.13482}{{
  arXiv:2210.13482~[cond-mat.stat-mech]}}.

\bibitem{Meneses:2018xpu}
S.~Meneses, J.~Penedones, S.~Rychkov, J.~M. Viana Parente~Lopes, and
  P.~Yvernay, {\slshape {A structural test for the conformal invariance of the
  critical 3d Ising model},}
  \href{http://dx.doi.org/10.1007/JHEP04(2019)115}{{\em JHEP} {\bfseries 04}
  (2019) 115}, \href{http://arxiv.org/abs/1802.02319}{{
  arXiv:1802.02319~[hep-th]}}.

\bibitem{Delamotte:2018fnz}
B.~Delamotte, M.~Tissier, and N.~Wschebor, {\slshape {Comment on `A structural
  test for the conformal invariance of the critical 3d Ising model' by S.
  Meneses, S. Rychkov, J. M. Viana Parente Lopes and P. Yvernay.
  arXiv:$1802$.$02319$},} \href{http://arxiv.org/abs/1802.07157}{{
  arXiv:1802.07157~[hep-th]}}.

\bibitem{Beisert:2003tq}
N.~Beisert, C.~Kristjansen, and M.~Staudacher, {\slshape {The dilatation
  operator of conformal $\mathcal N=4$ super Yang-Mills theory},}
  \href{http://dx.doi.org/10.1016/S0550-3213(03)00406-1}{{\em Nucl. Phys. B}
  {\bfseries 664} (2003) 131--184},
  \href{http://arxiv.org/abs/hep-th/0303060}{{ arXiv:hep-th/0303060}}.

\bibitem{Beisert:2003jj}
N.~Beisert, {\slshape {The complete one loop dilatation operator of $\mathcal
  N=4$ super Yang-Mills theory},}
  \href{http://dx.doi.org/10.1016/j.nuclphysb.2003.10.019}{{\em Nucl. Phys. B}
  {\bfseries 676} (2004) 3--42}, \href{http://arxiv.org/abs/hep-th/0307015}{{
  arXiv:hep-th/0307015}}.

\bibitem{Alonso:2016oah}
R.~Alonso, E.~E. Jenkins, and A.~V. Manohar, {\slshape {Geometry of the scalar
  sector},} \href{http://dx.doi.org/10.1007/JHEP08(2016)101}{{\em JHEP}
  {\bfseries 08} (2016) 101}, \href{http://arxiv.org/abs/1605.03602}{{
  arXiv:1605.03602~[hep-ph]}}.

\bibitem{Helset:2022tlf}
A.~Helset, E.~E. Jenkins, and A.~V. Manohar, {\slshape {Geometry in scattering
  amplitudes},} \href{http://dx.doi.org/10.1103/PhysRevD.106.116018}{{\em Phys.
  Rev. D} {\bfseries 106} (2022) 116018},
  \href{http://arxiv.org/abs/2210.08000}{{ arXiv:2210.08000~[hep-ph]}}.

\bibitem{Mihaila:2013dta}
L.~Mihaila, {\slshape {Three-loop gauge beta function in non-simple gauge
  groups},} \href{http://dx.doi.org/10.22323/1.197.0060}{{\em PoS} {\bfseries
  RADCOR2013} (2013) 060}.

\bibitem{Poole:2019kcm}
C.~Poole and A.~E. Thomsen, {\slshape {Constraints on $3$- and $4$-loop
  $\beta$-functions in a general four-dimensional Quantum Field Theory},}
  \href{http://dx.doi.org/10.1007/JHEP09(2019)055}{{\em JHEP} {\bfseries 09}
  (2019) 055}, \href{http://arxiv.org/abs/1906.04625}{{
  arXiv:1906.04625~[hep-th]}}.

\bibitem{Steudtner:2020tzo}
T.~Steudtner, {\slshape {General scalar renormalisation group equations at
  three-loop order},} \href{http://dx.doi.org/10.1007/JHEP12(2020)012}{{\em
  JHEP} {\bfseries 12} (2020) 012}, \href{http://arxiv.org/abs/2007.06591}{{
  arXiv:2007.06591~[hep-th]}}.

\bibitem{Steudtner:2021fzs}
T.~Steudtner, {\slshape {Towards general scalar-Yukawa renormalisation group
  equations at three-loop order},}
  \href{http://dx.doi.org/10.1007/JHEP05(2021)060}{{\em JHEP} {\bfseries 05}
  (2021) 060}, \href{http://arxiv.org/abs/2101.05823}{{
  arXiv:2101.05823~[hep-th]}}.

\bibitem{Bednyakov:2021qxa}
A.~Bednyakov and A.~Pikelner, {\slshape {Four-loop gauge and three-loop Yukawa
  beta functions in a general renormalizable theory},}
  \href{http://dx.doi.org/10.1103/PhysRevLett.127.041801}{{\em Phys. Rev.
  Lett.} {\bfseries 127} (2021) 041801},
  \href{http://arxiv.org/abs/2105.09918}{{ arXiv:2105.09918~[hep-ph]}}.

\bibitem{Davies:2021mnc}
J.~Davies, F.~Herren, and A.~E. Thomsen, {\slshape {General
  gauge-Yukawa-quartic $\beta$-functions at 4-3-2-loop order},}
  \href{http://dx.doi.org/10.1007/JHEP01(2022)051}{{\em JHEP} {\bfseries 01}
  (2022) 051}, \href{http://arxiv.org/abs/2110.05496}{{
  arXiv:2110.05496~[hep-ph]}}.

\bibitem{Steudtner:2024teg}
T.~Steudtner and A.~E. Thomsen, {\slshape {General quartic $\beta$-function at
  three loops},} \href{http://dx.doi.org/10.1007/JHEP10(2024)163}{{\em JHEP}
  {\bfseries 10} (2024) 163}, \href{http://arxiv.org/abs/2408.05267}{{
  arXiv:2408.05267~[hep-ph]}}.

\bibitem{Jack:1982hf}
I.~Jack and H.~Osborn, {\slshape {Two loop background field calculations for
  arbitrary background fields},}
  \href{http://dx.doi.org/10.1016/0550-3213(82)90212-7}{{\em Nucl. Phys. B}
  {\bfseries 207} (1982) 474--504}.

\bibitem{Jack:1982sr}
I.~Jack and H.~Osborn, {\slshape {General two loop beta functions for gauge
  theories with arbitrary scalar fields},}
  \href{http://dx.doi.org/10.1088/0305-4470/16/5/026}{{\em J. Phys. A}
  {\bfseries 16} (1983) 1101}.

\bibitem{Jack:1983sk}
I.~Jack and H.~Osborn, {\slshape {Background field calculations in curved
  space-time $1$. General formalism and application to scalar fields},}
  \href{http://dx.doi.org/10.1016/0550-3213(84)90067-1}{{\em Nucl. Phys. B}
  {\bfseries 234} (1984) 331--364}.

\bibitem{Jack:1984vj}
I.~Jack and H.~Osborn, {\slshape {General background field calculations with
  fermion fields},} \href{http://dx.doi.org/10.1016/0550-3213(85)90088-4}{{\em
  Nucl. Phys. B} {\bfseries 249} (1985) 472--506}.

\bibitem{Fonseca:2025zjb}
R.~M. Fonseca, P.~Olgoso, and J.~Santiago, {\slshape {Renormalization of
  general Effective Field Theories: formalism and renormalization of bosonic
  operators},} \href{http://dx.doi.org/10.1007/JHEP07(2025)135}{{\em JHEP}
  {\bfseries 07} (2025) 135}, \href{http://arxiv.org/abs/2501.13185}{{
  arXiv:2501.13185~[hep-ph]}}.

\bibitem{Aebischer:2025zxg}
J.~Aebischer, L.~C. Bresciani, and N.~Selimovic, {\slshape {Anomalous dimension
  of a general effective gauge theory. Part I. Bosonic sector},}
  \href{http://dx.doi.org/10.1007/JHEP08(2025)209}{{\em JHEP} {\bfseries 08}
  (2025) 209}, \href{http://arxiv.org/abs/2502.14030}{{
  arXiv:2502.14030~[hep-ph]}}.

\bibitem{Misiak:2025xzq}
M.~Misiak and I.~Na{\l}{\k{e}}cz, {\slshape {One-loop renormalization group
  equations in generic effective field theories. Part I. Bosonic operators},}
  \href{http://dx.doi.org/10.1007/JHEP06(2025)210}{{\em JHEP} {\bfseries 06}
  (2025) 210}, \href{http://arxiv.org/abs/2501.17134}{{
  arXiv:2501.17134~[hep-ph]}}.

\bibitem{Caswell:1974gg}
W.~E. Caswell, {\slshape {Asymptotic behavior of nonabelian gauge theories to
  two-loop order},} \href{http://dx.doi.org/10.1103/PhysRevLett.33.244}{{\em
  Phys. Rev. Lett.} {\bfseries 33} (1974) 244}.

\bibitem{Banks:1981nn}
T.~Banks and A.~Zaks, {\slshape {On the phase structure of vector-like gauge
  theories with massless fermions},}
  \href{http://dx.doi.org/10.1016/0550-3213(82)90035-9}{{\em Nucl. Phys. B}
  {\bfseries 196} (1982) 189--204}.

\bibitem{Hansen:2017pwe}
F.~F. Hansen, T.~Janowski, K.~Lang{\ae}ble, R.~B. Mann, F.~Sannino, T.~G.
  Steele, and Z.-W. Wang, {\slshape {Phase structure of complete asymptotically
  free $SU(N_c)$ theories with quarks and scalar quarks},}
  \href{http://dx.doi.org/10.1103/PhysRevD.97.065014}{{\em Phys. Rev. D}
  {\bfseries 97} (2018) 065014}, \href{http://arxiv.org/abs/1706.06402}{{
  arXiv:1706.06402~[hep-ph]}}.

\bibitem{Fei:2016sgs}
L.~Fei, S.~Giombi, I.~R. Klebanov, and G.~Tarnopolsky, {\slshape {Yukawa CFTs
  and emergent supersymmetry},}
  \href{http://dx.doi.org/10.1093/ptep/ptw120}{{\em PTEP} {\bfseries 2016}
  (2016) 12C105}, \href{http://arxiv.org/abs/1607.05316}{{
  arXiv:1607.05316~[hep-th]}}.

\bibitem{Gracey:2016mio}
J.~A. Gracey, T.~Luthe, and Y.~Schroder, {\slshape {Four loop renormalization
  of the Gross-Neveu model},}
  \href{http://dx.doi.org/10.1103/PhysRevD.94.125028}{{\em Phys. Rev. D}
  {\bfseries 94} (2016) 125028}, \href{http://arxiv.org/abs/1609.05071}{{
  arXiv:1609.05071~[hep-th]}}.

\bibitem{Ihrig:2018hho}
B.~Ihrig, L.~N. Mihaila, and M.~M. Scherer, {\slshape {Critical behavior of
  Dirac fermions from perturbative renormalization},}
  \href{http://dx.doi.org/10.1103/PhysRevB.98.125109}{{\em Phys. Rev. B}
  {\bfseries 98} (2018) 125109}, \href{http://arxiv.org/abs/1806.04977}{{
  arXiv:1806.04977~[cond-mat.str-el]}}.

\end{thebibliography}\endgroup

\clearpage
\newpage

\widetext
\begin{center}
\section*{Supplemental Material}
\end{center}
\vspace{-0.2cm}
%%%%%%%%%% Prefix a "S" to all equations, figures, tables and reset the counter %%%%%%%%%%
\setcounter{equation}{0}
\setcounter{figure}{0}
\setcounter{table}{0}
\setcounter{page}{1}
\makeatletter
\renewcommand{\theequation}{S\arabic{equation}}
\renewcommand{\thefigure}{S\arabic{figure}}
\renewcommand{\thepage}{S\arabic{page}}
% \renewcommand{\bibnumfmt}[1]{[S#1]}
% \renewcommand{\citenumfont}[1]{S#1}
%%%%%%%%%% Prefix a "S" to all equations, figures, tables and reset the counter %%%%%%%%%%
\subsection*{Lagrangian}
The Lagrangian used in our computations is
\begin{equation}\label{eq:gentheory}
    \mathcal{L} = \frac12 \partial_\mu \phi^a \partial^\mu \phi^a 
    - \frac{1}{4!} \lambda^{abcd} \phi^a \phi^b \phi^c \phi^d 
    + \mathcal{L}_\text{rel}
    + \mathcal{L}_\text{spin-0}
    + u_{\mu}\mathcal{L}_\text{spin-1}^\mu 
    + v_{(\mu\nu)} \mathcal{L}_{\text{spin-2}}^{(\mu\nu)}
    + w_{[\mu\nu]} \mathcal{L}_{\text{spin-2}}^{[\mu\nu]}
    \,,
\end{equation}
where $u_\mu$, $v_{(\mu\nu)}$ and $w_{[\mu\nu]}$ are reference Lorentz tensors; $v$ is traceless symmetric and $w$ is antisymmetric. 
The composite operators are defined by
\begin{align}
    \mathcal{L}_\text{rel} &= 
        % -\Lambda
        - t_a \phi^a 
        - \frac{m^2_{(ab)}}{2}\,\phi^a\phi^b
        - \frac{h_{(abc)}}{3!}\,\phi^a\phi^b\phi^c\,,
    \end{align}
    and 
    {\allowdisplaybreaks
    \begin{align}
    \mathcal{L}_\text{spin-0} &=
    \frac{c_{(abcde)}^{(5,0)}}{5!} \, 
        \phi^a \phi^b \phi^c \phi^d \phi^e
    +\frac{c_{(abcdef)}^{(6,0)}}{6!}\,
        \phi^a\phi^b\phi^c\phi^d\phi^e \phi^f
    -\frac{c_{((ab),(cd))}^{(6,0)}}{4}\,
        \phi^a\phi^b\partial_\mu\phi^c\partial^\mu \phi^d\,,\nonumber\\[4mm]
    \mathcal{L}_\text{spin-1}^\mu &=
    c_{[ab]}^{(3,1)}\,\phi^a \partial^\mu \phi^b
    +\frac{c_{(ab)c}^{(4,1)}}{2} \phi^a \phi^b \partial^\mu \phi^c
    +\frac{c_{(abc)d}^{(5,1)}}{3!}  \phi^a \phi^b \phi^c \partial^\mu \phi^d\nonumber
    +c_{[ab]}^{(5,1)} \left(
      3\,\partial^\mu \phi^{a} \partial^2 \phi^{b}
      -\phi^{a} \partial^\mu \partial^2 \phi^{b}
   \right)
    \\&\qquad 
    +\frac{c_{(abcd)e}^{(6,1)}}{4!} \phi^a\phi^b\phi^c\phi^d \partial^\mu\phi^e\,,\nonumber\\[4mm]
    \mathcal{L}_\text{spin-2}^{(\mu\nu)} &=
    c^{(4,2)}_{(ab)}
    \left(\phi^a\partial^\mu\partial^\nu \phi^b -2\,\partial^\mu\phi^a\partial^\nu\phi^b\right)
    + 2 \, c_{a(bc)}^{(5,2)} \left( 
    \phi^a\phi^b\partial^\mu\partial^\nu\phi^c 
    -2 \, \phi^a \partial^\mu\phi^b \partial^\nu \phi^c 
    \right)\nonumber\\&\qquad 
    + c_{(ab)(cd)}^{(6,2)}
    \left(
    \phi^a\phi^b\phi^c
    \partial^\mu\partial^\nu\phi^d
    -2 \, \phi^a\phi^b
    \partial^\mu\phi^c\partial^\nu\phi^d
    \right)
    +c_{(ab)}^{(6,2)}
      \partial^\mu \partial^\nu \phi^{a} 
      \partial^2 \phi^{b}\,,
      \nonumber\\[4mm]
    \mathcal{L}_\text{spin-2}^{[\mu\nu]} &=
    -c_{[abc]}^{(5,\{1,1\})} \, \phi^a \partial^\mu \phi^b \partial^\nu \phi^c 
    -c_{a[bcd]}^{(6,\{1,1\})}\,
    \phi^a\phi^b
    \partial^\mu\phi^c\partial^\nu\phi^{d}\,,
    \label{eq:LagrsGenTheory}
\end{align}
where we indicated the relabeling symmetries of the tensors using brackets. For example, $c^{(6,0)}_{((ab),(cd))}$ is symmetric in its first two indices and in its last two indices, as well as under the exchange of the two pairs; and $c^{(6,\{1,1\})}_{a[bcd]}$ is antisymmetric in its last three indices.
We list the constraints that arise from the primary condition in the explanatory files in the \texttt{GitHub} repository \href{https://github.com/jasperrn/EFT-RGE}{https://github.com/jasperrn/EFT-RGE}. All calculations have been performed with the primary condition imposed.

Note that in \eqref{eq:LagrsGenTheory} we have kept the redundant operators with coefficients $c^{(5,1)}_{ab}$ and $c^{(6,2)}_{ab}$. This is because the field redefinitions to remove these operators, such as~$\phi^a \to  \phi^a + 4\,c^{(5,1)}_{ab} u_\mu \, \partial^\mu \phi^b$, 
result in an infinite wavefunction normalization factor, which complicates the LSZ reduction formula.

\subsection*{$R^*_{\text{ME}}$ examples}
Here we give an example on how to compute the UV counterterm with $R^*_{\text{ME}}$. We start with a simple 1-loop counterterm for a $4$-point graph:
\begin{align}
\mathcal{Z} 
\left(
\oneloop
\quad\right)
&=-K_\eps \widetilde T_{p_3}^{(0)} \bar R
\left(
\oneloop
\quad\right)
=-K_\eps \widetilde T_{p_3}^{(0)} 
\left(
\oneloop
\quad\right)\nonumber\\
% &= -K_\eps \widetilde T_{p_3}^{(0)} 
% \left(
% \oneloop
% \quad\right)
&= -K_\eps
\left(
\oneloopa
\quad\right)
=-\frac{1}{\eps}\,.
\end{align}
In the first step, we simply substitute the definition for $\mathcal{Z}$ in the $R^*_{\text{ME}}$ approach. We have to make a choice on the external momentum to expand. Any choice will lead to the same result, and we can choose $p_3$. In the second step, we perform the operation $\bar R$. 
There is nothing to do since there are no sub-divergences at one loop. We then act with $\widetilde T_{p_3}^{(0)}$. 
For this simple example, this just leads to setting $p_3$ to zero; there are no other AI subgraphs which can lead non-vanishing results. Finally, we perform the integration and act with the pole operator $K_\eps$.

Let us now consider a two-loop example.
\begin{align}
\mathcal{Z} 
\Bigg(&
\twoloop
\quad\Bigg)
=-K_\eps \widetilde T_{p_3}^{(0)} \bar R
\Bigg(
\twoloop
\quad\Bigg)\nonumber\\
&\qquad=-K_\eps \widetilde T_{p_3}^{(0)} 
\Bigg(
\twoloop
+
\mathcal{Z} 
\Bigg(\;\;\gsubb\Bigg) \cdot \gsuba
\quad\Bigg)\nonumber\\
&\qquad=-K_\eps 
\Bigg(\widetilde T_{p_3}^{(0)} 
\Bigg[
\twoloop\quad
\Bigg]
- \frac{1}{\eps}
\widetilde T_{p_3}^{(0)} 
\Bigg[ \gsuba
\quad\Bigg]\Bigg)\\
&\qquad=-K_\eps 
\Bigg(
\Bigg[ \ 
\twoloopa
+
\gsuba \cdot \gsubb
\Bigg]
- \frac{1}{\eps}
 \gsuba
\quad\Bigg)\nonumber\\
&\qquad=-K_\eps\Bigg(\bigg[\Big( -\frac{1}{2\eps^2}
+\frac{-3+2\log (p_1^2)}{2\eps} \Big)  
+\Big(\frac{1}{\eps}+2-\log(p_3^2)\Big)\Big(\frac{1}{\eps}+2-\log(p_1^2)\Big)
\bigg] \nonumber\\          
&\qquad\qquad \qquad -\frac{1}{\eps}\Big(\frac{1}{\eps}+2-\log(p_3^2)\Big)+O(\eps^0)\Bigg)=-\frac{1}{2\eps}+\frac{1}{2\eps^2}\,.\nonumber
\end{align}
Here we proceeded as before, just that this time there is a log-divergent UV sub-divergence, which obtains a counterterm in the second line after acting with $\bar R$. Note that the insertion symbol ``$*$'' is reduced to an ordinary product ``$\cdot$'' for log-divergent subgraphs. In the third line, we evaluate this counterterm and then continue with the expansion in $p_3$. For the two-loop graph two asymptotically irreducible subgraphs contribute, the full graph and the bubble subgraph, which connects to both hard momenta $p_1$ and $p_2$. 
For the contracted one-loop graph there is no expansion to be done as it is already homogeneous in $p_3$, i.e.~only the vertex connecting $p_1, p_2$ is AI. 
When evaluating the remaining integrals all logarithms cancel, yielding the well-known result for this graph.     
\end{document}